\documentclass[12pt]{iopart}

\usepackage{iopams}  

\usepackage{tikz}

\usepackage{graphicx}

\usetikzlibrary{shapes.arrows}

\usetikzlibrary{arrows,decorations.pathmorphing}

\usetikzlibrary{decorations.pathreplacing,angles,quotes}

\pgfdeclareradialshading[tikz@ball]{ball}{\pgfqpoint{0bp}{0bp}}{%
 color(0bp)=(tikz@ball!0!white);
 color(7bp)=(tikz@ball!0!white);
 color(15bp)=(tikz@ball!70!black);
 color(20bp)=(black!70);
 color(30bp)=(black!70)}

\pgfdeclarefunctionalshading{sphere}{\pgfpoint{-25bp}{-25bp}}{\pgfpoint{25bp}{25bp}}{}{
25 div exch
25 div exch
2 copy 
dup mul exch
dup mul add
1.0 sub
0.3 dup mul
-0.5 dup mul add
1.0 sub
mul abs sqrt
exch 0.3 mul add
exch -0.5 mul add
dup abs add 2.0 div 
.6 mul .4 add
dup
4.5
}

\def\halb{{\frac{1} {2}}}
\def\d{\delta}
\def\e{\epsilon}
\def\g{\gamma}

\begin{document}

\title[Poincar\'e and Einstein on Mass-Energy Equivalence]{Poincar\'e and Einstein on Mass-Energy Equivalence: A Modern Perspective on their 1900 and 1905 Papers}

\author{Patrick Moylan}

\address{Physics Department, The Pennsylvania State University,
Abington College, Abington, PA 19001, USA}
\ead{pjm11@psu.edu}
\vspace{10pt}
\begin{indented}
\item[]
\end{indented}

\begin{abstract} 
Both Poincar\'e in his 1900 Festschrift paper \cite{Poincare} and Einstein in his 1905 \textsl{Annalen der Physik} article \cite{Einstein} were led to $E=mc^2$ by considering electromagnetic processes taking place in vacuo.  Poincar\'e's treatment is based on a generalization of the law of conservation of momentum to include radiation.  Einstein's analysis relies solely on energy conservation and the relativity principle together with certain assumptions, which have served as the source of criticism of the paper beginning with Max Planck in 1907.  We show that these objections raised by Planck and others can be traced back to Einstein's failure to make use of momentum considerations.   Relevance of our findings to a proper understanding of Ives' criticism of Einstein's paper is pointed out.

\end{abstract}

\section{Introduction}

In [Eur. J. Phys. {\bf 25} (2014) 123-126],  J. H. Field provides a striking vindication of Einstein's 1905 paper on $E=mc^2$ undoing more than 100 years of criticism which started with Max Planck \cite{Planck}.   Field's article \cite{Field} serves as an attempt to refute all reputable claims as to possible shortcomings in Einstein's  paper.  We wish to show that such is far from the case.   In this paper, we restrict ourselves to refuting his attack on the Ives criticism of Einstein's 1905 paper. 

Planck's objection to Einstein's article seems to have gone unoticed, at least as far as written criticism was concerned, until many years later, when Herbert E Ives elaboration on  Planck's remarks appeared in the Journal of the Optical Society of America \cite{Ives}.  This resulted in subsequent criticism of the paper, most notably, by Max Jammer, the renown physicist and philosopher of science \cite{Jammer}, and the well-known Einstein biographer, Arthur I. Miller \cite{Miller}. Additional investigations into the paper's shortcomings include those of reputable physicists such as Arziel\'es \cite{Arzieles}.         

Einstein advocates have replied with counterclaims, the most notable of which  being that of Stachel and Torretti \cite{Stachel} published in the American Journal of Physics in 1982.  Overturning the judgment of Planck, Ives and others, it is their contention that ``the paper [Einstein's] was basically sound" \cite{Stachel}.   Stachel and Torretti's article is sometimes quoted as being the definitive judgment on the subject regarding the Ives criticism. Even Jammer in his 2009 revision \cite{Jammer2} of his book  \textit{Concepts of Mass}, first published in 1961, drastically moderated his original view presented in  the 1961 first edition \cite{Jammer}. He attributes his change in stance on the the matter to the Stachel and Torretti paper.      

In 2009 Ohanian \cite{Ohanian}  put a new twist on things.   Although he lent his support to Stachel and Torretti's refutation of the circular reasoning mistake argued by Ives \cite{Ives} and Jammer \cite{Jammer}, he did not agree that they were right; instead, he turned the argument around by insisting that the problem lies in the fact that the assumptions of Einstein, on which Planck and Ives based their criticisms, may  in fact not always be true, claiming the ``defect in Einstein's argument is not  a \textit{petitio principii}, but a \textit{non sequitur}" \cite{Ohanian}.  Ohanian's viewpoint has not gone without  counterattacks from Einstein advocates, especially by Mermin \cite{Mermin}, \cite{Ohanian2}, but also by Field in the Eur. J. Phys. paper referenced above \cite{Field}.   Finally, Hecht \cite{Hecht}, after making a very thorough analysis of all of Einstein's attempts to derive mass-energy equivalence,  was led to the conclusion that it may never be possible to give a completly satisfactory explanation of $E=mc^2$, essentially because we would have to take into account all forms of energy and matter, local and nonlocal, which contribute to the inertia of a body; specifically, according to Hecht: ``A proof of either $E_0=mc^2$ or   $E_R=mc^2$ is therefore inextricably linked to a determination of the fundamental nature of mass and is beyond the purview of the special theory."

An outline of the paper is as follows.  In Sections 2 and 3 we present Poincar\'e's and Einstein's treatments of mass-energy equivalence.  Section 4 contains a further development of Poincar\'e's approach to mass-energy equivalence along lines similar to Einstein's treatment. Our analysis in Section 4 shows the necessity for considering radiation momentum and momentum conservation in order to get at Poincar\'e's result on mass-energy equivalence, energy considerations alone do not suffice.  In Section 5  we argue that Einstein's approach based solely on energy considerations is either ambiguous or, by declaring the energies of the two light waves to be the same in the rest frame of the body, he must implicitly have made use of momentum conservation and momentum of radiation, which is what Stachel and Toretti \cite{Stachel} and Field \cite{Field} assume.  Without using momentum conservation Einstein has no right to single out the rest frame as the frame where the energies of the two light pulses are equal; he can only assert that the two energies are equal in some inertial frame, not necessarily in the rest frame of the body.  The relativity principle demands this be so. This leads to ambiguity in interpretation of the kinetic energy, which is the main point of the Ives' criticism.  Finally in Section 6 we show how a correct interpretation of the logic of the Ives' criticism of Einstein's paper presupposes an understanding of the analysis presented in Section 5.  In particular, a proper understanding of the logic of the Ives criticism can only be understood in terms of Einstein's failure to take into account momentum conservation.  In order be as clear as possible, we consider a special case of Einstein's emission process, namely, the case where the particle ceases to exist after the emission of radiation, an example of which is neutral pion decay into two photons.  The main results of the paper are Theorems 1 and 2. Theorem 1 contains the statement regarding the logical fallacy of Einstein's analysis as first explained by Ives \cite{Ives}.  Theorem 2 goes a little further.  It states that, \underline{without using momentum conservation}, equality of the energies of the two light waves in the rest frame of the body holds true if and only if $\d =0$, which, combined with Theorem 1, implies that without momentum conservation \textit{equality of the energies of the two light waves in the rest frame of the body holds true if and only $E_0=m_0 c^2$}.  In the conclusions we analyze the attacks on the Ives criticism, in particular those of Ref. \cite{Field} and \cite{Stachel}, in light of the Theorems and show that they are, from our way of looking at things, unjustified.

\section{Poincar\'e on Mass-Energy Equivalence} 
In the summer of 1900, Poincar\'e wrote a long memoir entitled “Lorentz’s theory and the
principle of reaction” for a volume devoted to the 25th anniversary of Lorentz' dissertation. Poincar\'e's memoir, frequently referred to as his Festschrift article, started off with an objection to  Lorentz’s electron theory in that it implies an intolerable violation of the reaction principle when applied solely to ponderable matter but, at the same time, praising Lorentz for having come up with the best theory possible.  

After this brief introduction, he then proceeds with an analysis of the Maxwell-Lorentz equations leading him to the following result: a system of charged particles of masses $m_i$  moving  with  velocities  ${\vec v}_i$ 
and  interacting  through  their electromagnetic fields  $\vec E$   and  $\vec B$ leads him to his Eq. (4), which in modern (M.K.S.) notation  is 
$$ \sum m_i{\vec v}_i + \frac{1}{c^2} \int  \vec S  dV = constant        \eqno(1)                  $$ 
where $$\vec S = \frac{1}{\mu_0} \vec E \times \vec B \eqno(2)$$
is the Poynting vector. From this it follows that the momentum of matter is, alone by itself, not conserved; rather,  we must also take into account the contribution from the momentum of the electromagnetic fields\footnote{Abraham cites Poincar\'e's 1900 article for the basic idea of
ascribing momentum to the electromagnetic field \cite{Abraham1}, \cite{Abraham2}.  Griffiths attributes it to Poynting \cite{Griffiths}. Surprisingly, Whittaker \cite{Whittaker} is silent on the matter. I have looked into the references given by Griffiths for justifiying his claim that it is due to Poynting.  Although Griffiths cites articles by Poynting on momentum of radiation written after 1905 \cite{Griffiths}, it turns out that there is no mention whatsoever about \textit{radiation momentum} in the Poynting's works pre-1905, and, for sure, nothing before 1900, the year of Poincar\'e's Festschrift article. In Poynting's 1883 paper \cite{Poynting}, where Poynting's important theorem is proved,  he writes only  of "movement of energy" across a surface.  In that article he does not endow his "movement of energy"  with any of the properties of momentum which Poincar\'e goes into great detail in establishing, such as, relating it to \textit{inertia of radiation} and generalizing center of gravity and momentum conservation to include radiation.  It does not seem surprising to me that Poynting in 1883 would not have entertain such ideas.  At that time, such notions would have been very strange and puzzling.  Poynting would not have yet known of Lorentz' works. He might have familiarized himself with Hertz' theory (1882), but that was a Galilean invariant theory of electromagnetism respecting the law of action and reaction as a law between ponderable material objects and thus prohibiting any generalization of the law of conservation of momentum to include momentum of radiation.  

I have invested some effort into researching the matter as to whom should go the credit for first having come up with momentum of radiation, using mostly first sources,  and restricting myself to times around the time of the  Ponyting's paper on the Poynting theorem (1883), since, it seems to me that radiation momentum could not have been dealt with in any serious way prior to Poynting's theorem. For sure, the notion is rooted in the   electromagnetic models of the electron which date back as least as far as 1881 when J.J. Thompson  argued that the backreaction of the magnetic field of a charged sphere would impede its motion and result in an apparent mass increase of the sphere \cite{Thomson}.  Shortly thereafter, Heaviside proved that the mass increase of a moving
sphere with uniform surface charge distribution was $m = (4/3)E_0/c^2$ where $E_0$ is the electromagnetic energy of a stationary sphere \cite{Heaviside}.  But these investigations did not consider the electromagnetic field as a separate entity, independent of the charges which produce it, and  they said nothing about how to incorporate electromagnetic energy into a consistent physical theory generalizing mechanical laws, such as the laws of conservation of energy and momentum.  Likewise, it is difficult to imagine that Poynting could have initiated such decisive steps. To do such required someone like Poincar\'e, who fully appreciated and was fully engaged in the creation of the revolution in physics that was taking place at that time, to assert that radiation carried inertia.  Indeed it does seem that Abraham is justified in attributing \textit{momentum of radiation} to Poincar\'e.  The authors of  Refs. \cite{Ives} and \cite{Janssen} also share the opinion that it originated with Poincar\'e.}   in order to uphold the law of conservation of momentum.  Specifically the momentum of the radiation is given by the second term on the left hand side of the above equation and its momentum density is 
$$ \vec g =  \frac{1}{\mu_0c } \vec E \times \vec B .\eqno(3)$$
The electromagnetic field associated with this electromagnetic momentum is described by the pair $(\vec E, \vec B)$ with $\vec E$ and $\vec B$ being the electric and magnetic fields, respectively and Poincar\'e explains how it can be viewed a fluid carrying momentum.  With a mathematician's proclivity for rigor, Poincar\'e cannot allow himself to assert that it may be viewed in all situations as a real fluid, so instead he calls it  ``fluide fictif."   Only when 
$$\int dV \rho \vec v \cdot \vec E =0 \eqno(4)$$ 
where $\vec v$ is the velocity of a material element and $\rho$ the charge density of the matter, can one consider it as a real fluid \cite{Poincare}.    

For a point charge moving with velocity $\vec v$ and a plane polarized electromagnetic wave propagating in a direction parallel to the direction of the velocity of the charge, $\vec v \cdot \vec E$ vanishes.  In this case, the above condition, Eq. (4),  is satisfied, and so it can viewed as a real fluid.  

Poincar\'e illustrates his findings with an example of such, i.e. an example satisfying Eq. (4), for which the radiation can be treated as real fluid.  Specifically, he considers a  device, which he calls a Hertzian exciter, that produces electromagnetic energy and radiates it in a particular direction.  According to Poincar\'e ``that device must
recoil just as a cannon does when it fires a projectile."   In Fig. 1.  we represent the recoiling device as a point charge of mass $M_0$ and the emitted radiation as a plane wave which is ``propagated along the $x$ axis (horizontal) in the positive direction"  opposite to the direction of recoil of the device. The original mass of the device before the emission of radiation is $M$ and the recoil velocity is $v$.  This example, according to Poincar\'e, suffices to establish that radiation posesses inertia, with the amount of inertia given by  $E/c^2$ where $E$ is the energy of the radiation.  

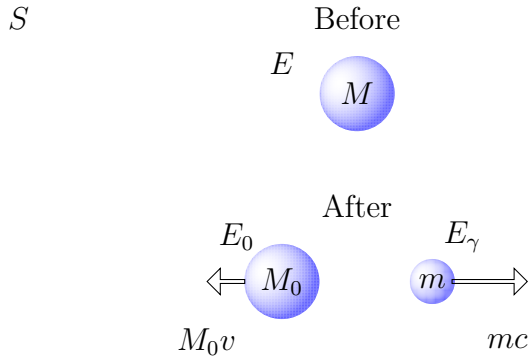
\begin{figure}[htb]
\begin{center}
\begin{tikzpicture}[
    media/.style={font={\footnotesize\sffamily}},
    wave/.style={
        decorate,decoration={snake,post length=1.4mm,amplitude=.6mm,
        segment length=2mm},thick},
    interface/.style={postaction={draw,decorate,decoration={border,angle=-45,
                    amplitude=0.3cm,segment length=2mm}}},scale=1]

\node at (4,0) {$~~~~~~~~~~~~~~~~~$};
 \shade[shading=sphere] (-1,0) circle [radius=.5cm];

 \shade[shading=sphere] (0,2.5) circle [radius=.5cm];

 \shade[shading=sphere] (1,0) circle [radius=.3cm];

\node at (-1.73,0) [single arrow,draw,inner sep=.3,minimum width=.35cm,minimum height=.5cm,single arrow head extend=0.1cm, rotate=180] {};

\node at (-4.5, 3.5) {$S$};

\node at (1.4,.6) {$E_\gamma$};
\node at (-1.6,.6) {$E_0$};

\node at (2.0,-.8) {$mc$};
\node at (-2.0,-.8) {$M_0v$};

\node at (0, 3.5) {$\textrm{Before}$};

\node at (0, 1) {$\textrm{After}$};

\node at (1.75,0) [ single arrow, draw,inner sep=.3,minimum width=.35cm,minimum height=1cm,single arrow head extend=0.1cm] {};

\node at (-1, 2.9) {$E$};

\node at (0, 2.5) {$M$};

\node at (1, 0) {$m$};
\node at (-1, 0) {$M_0$};

\end{tikzpicture}
\caption{ Poincar\'e's recoil process.  A body (Hertzian exciter) initially at rest and of mass $M$ emits a light pulse of energy $E_\g$.  After the emission, the body ``must recoil just as a cannon does \cite{Poincare}."  The mass of the body after the emission is $M_0$ and its velocity is $\vec v$. The mass $m$ of the light pulse is determined by Eq. (6). }  
\end{center}
\end{figure}

We see this as follows.  Returning momentarily to the general situation, i.e. where $\vec v \cdot \vec E$ does not necessarily vanish, so that the fluid cannot necessarily be considered as a real fluid, and the matter distribution is arbitrary, Poincar\'e writes in Ref. \cite{Poincare}:
``We shall let $M_0$ represent the total mass of the matter, ${\vec R}_0$ for the coordinates of its center of gravity, $m$ to represent the total mass of the fictional fluid, ${\vec r}$ for the coordinates of its center of gravity, $M$ for the total mass of the system (matter plus fictional fluid), and ${\vec R}$ for its center of gravity.  We shall then have
$$ M= M_0 + m \; ,  \; M   {\vec R} = M_0 {\vec R}_0 +m {\vec r} \eqno(5)$$
$$  \frac{1}{c^2}\int {{\vec {r'}}} \, {u_{E.M.}}\, d{ V'} = m  {\vec r} \quad "\eqno(6)$$ 
(cf. the equations just above the second of Poincar\'e's two equation (3)'s in Ref. \cite{Poincare}).
Earlier in the paper, Poincar\'e defined  $u_{\textrm{E.M.}}$ as $$``u_{\textrm{E.M.} }= \frac{\e_0}2|{\vec E}|^2 + \frac{1}{2\mu_0}  |{\vec B}|^2 ,"$$ where $u_{\textrm{E.M.}}$ is the energy density of the electromagnetic fluid (cf. the equation immediately above the first of his two equation (3)'s in Ref. \cite{Poincare}).  This equation and Eq. (6) uniquely define the mass $m$ of  Poincar\'e's {fluid} in terms of $\vec E$ and $\vec B$.  

Now let us return to the case considered in Fig. 1 where we treat the recoiling body as a point charge and the electromagnetic (plane wave) pulse as a \textit{real fluid} propagating in a direction opposite  to the direction of motion of the recoiling body.  With $E_\g$ being the total energy of the fluid and $$u_{\textrm{E.M.}}= E_\g \delta^3({\vec r'}- {\vec r}),\eqno(7)$$ 
so that the energy of the wave pulse is localized at a point, the integral on the left hand side of Eq. (6)  is $E_\g$  times ${\vec r}$, which gives $$E_\g = mc^2 ,\eqno(8)$$
relating the energy and the mass of the wave pulse.

Furthermore, since $|\vec E  |= c|\vec B|$ for superpositions of plane harmonic electromagnetic waves, we have $|\vec S| = \frac{1}{\mu_0 c} |\vec E|^2 = \e_0 c|\vec E|^2 = c u_{E.M.}$  
so that the magnitude of the momentum density, Eq. (3), is $|\vec g| =\frac{u_{E.M.}}c$.  The momentum of the wave pulse $\vec p_\g$ is obtained by differentiating the second of the two equations in Eq. (5), and identifying it with the last term of the resultant equation.  Thus 
$$ mc = m \left\vert \frac{d \vec r}{dt} \right\vert  = p_\g  = \int |\vec g| \, dV = \frac{E_\g}c\eqno(9)$$
where $m$ is  the mass of the radiation, uniquely specified by Poincar\'e in terms of the electromagnetic field as explained above.  \\

Finally, Poincar\'e describes how all of this is to be understood in ``everyday language" as follows \cite{Poincare}:    \begin{quote} consider ``$\dots$  a light pulse emitted from a Hertzian exciter and causing the emitter to suffer a recoil just as a canon does. \dots It is easy to evaluate that recoil quantitatively. If the device has a mass of 1 kg and if it emits three million joules in one direction with the velocity of light, the speed of the recoil is 1 cm/sec." 
\end{quote}

\section{Einstein on Mass-Energy Equivalence}

Einstein's first paper on mass-energy equivalence,  most probably written sometime in the summer of 1905,  was a short three page article intended to describe what he considered to be a very important application of the results contained in his first relativity paper \cite{Einstein1}, in particular, as an application of his relativistic Doppler formula for light.  A summary of the paper is as follows.  Einstein considers two plane light waves simultaneously emitted in opposite directions by a body which is assumed stationary both before and after the emission of the light waves.   We let $E_0$ be the energy of the body in  the rest frame $S$ of the body  before the emission of electromagnetic energy and we let $E_1$ be the energy of the body in $S$ after the emission of light.  This situation is shown in Fig. 2.  We shall refer to the two plane light waves shown in Fig. 2 as light pulses or radiation pulses (or sometimes even photons as in Section 6). The energy of the light pulse moving to the right is $E_{\g_+}$ and the energy of the other one moving to the left is $E_{\g_-}$.   Without providing justification Einstein declares the energies $E_{\g_+}$ and $E_{\g_-}$ equal to each other. In Einstein's notation, $E_{\g_{+}}= E_{\g_-} = \halb L$ where $L =E_{\g_+} + E_{\g_-}$ is their total energy as measured in the inertial frame $S$.

\begin{figure}[htb]
\begin{center}
\begin{tikzpicture}[
    media/.style={font={\footnotesize\sffamily}},
    wave/.style={decorate,decoration={snake,post length=1.4mm,amplitude=.6mm,
        segment length=2mm},thick},
    interface/.style={postaction={draw,decorate,decoration={border,angle=-45,
                    amplitude=0.3cm,segment length=2mm}}},scale=.77]

    \draw[->,wave]  
(-180:.4cm)--(-180:3.2cm)node[left]{$~$};

\node at (-4.5, 3) {$S$};
    \draw[->,wave]
       (0:.4cm)--(0:3.2cm)node[right]{};
\node at (-2,+.5) {$\gamma_-$};
\node at (2.5,-.8) {$E_{\gamma_+}$};
\node at (-2.5,-.8) {$E_{\gamma_-}$};

\node at (2,+.5) {$\gamma_+$};

\node at (0, 1.5) {$E_0$};

\node at (0, 0) {$E_1$};
\draw[line width=1pt](0,1.5)circle(.4cm);
\node at (5.6, 2) {$~~~~~~~$};
 \draw[line width=1pt](0,0)circle(.4cm);

\end{tikzpicture}

\caption{\footnotesize{Emission of two plane light waves (light pulses) from a body as viewed in the rest frame $S$ of the body, which Einstein assumes remains at rest after the emission of radiation.  The light pulse with energy $E_{\g_+}$ moves to the right and the other one with energy $E_{\g_-}$ moves to the left. The total energy of the body before the emission of light is $E_0$ and the energy of the body after the emission of the two light pulses is $E_1$.  \textit{No justification for assuming that $E_{\g_+} = E_{\g_-}$ is provided by Einstein.}}}
\end{center}
\end{figure}
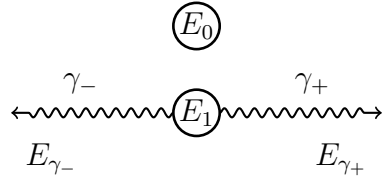

Einstein then considers the same process as viewed in another inertial reference frame $S'$ moving with speed $v$ ($< c$ = speed of light in vacuo) relative to $S$.  The situation in the frame $S'$ is shown in Fig. 3. To an observer in $S'$ the body moves with constant speed $v$ both before and after the emission of the two light pulses, i.e. the two plane light waves.      We let   $H_0$ be the total energy of the body in  the frame  $S'$ before the emission of electromagnetic energy.  We let  $H_1$ be the total energy of the body in $S'$ after the emission of the radiation.  Furthermore we let $K_0$ be the relativistic kinetic energy of the body in  the frame  $S'$ before the emission of electromagnetic energy and we let  $K_1$ be the relativistic kinetic energy of the body in $S'$ after the emission of the radiation.  The energies of the two light pulses in $S'$ are denoted by  $E'_{\g_+}$ and  $E'_{\g_-}$, and the total radiation energy in $S'$ is $L'=  E'_{\g_+} + E'_{\g_-}$.

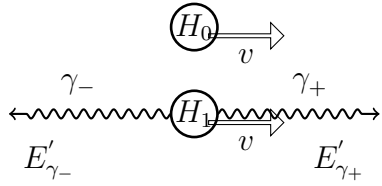
\begin{figure}
\begin{center}
\begin{tikzpicture}[
    media/.style={font={\footnotesize\sffamily}},
 wave/.style={decorate,decoration={snake,post length=1.4mm,amplitude=.6mm,
        segment length=2mm},thick},
    interface/.style={postaction={draw,decorate,decoration={border,angle=-45,
                    amplitude=0.3cm,segment length=2mm}}},
   scale=.77]

    \draw[->,wave]  
(-180:.4cm)--(-180:3.2cm)node[left]{$~$};

\node at (.88,-.15) [single arrow,draw,inner sep=.3,minimum width=.35cm,minimum height=1cm,single arrow head extend=0.1cm] {};
\node at (.88,1.35) [single arrow, draw,inner sep=.3,minimum width=.35cm,minimum height=1cm,single arrow head extend=0.1cm] {};
\node at (-4.5, 3) {$S'$};
    \draw[line width=1pt](0,1.5)circle(.4cm);

\node at (.9, -.5) {$v$};
   
    \draw[->,wave]
       (0:.4cm)--(0:3.2cm)node[right]{};

\node at (-2,+.5) {$\gamma_{-}$};

\node at (.9, 1.0) {$v$};
\node at (0, 1.5) {$H_0$};

\node at (0, 0) {$H_1$};

\node at (2.5,-.8) {$E_{\gamma_+}^{'}$};
\node at (-2.5,-.8) {$E_{\gamma_-}^{'}$};

\node at (2,+.5) {$\gamma_+^{}$};
 
    \draw[line width=1pt](0,0)circle(.4cm);

\end{tikzpicture}
\caption{\footnotesize{Emission of two plane light pulses from a body as viewed in the inertial frame $S'$.   The light pulse moving to the right now has energy $E'_{\g_{+}}$ and the other one has energy $E'_{\g_{-}}$. In $S'$ the total energy of the body before the emission of light is $H_0$ and the energy of the body after the emission of the two light pulses is $H_1$.  \textrm{In $S''$  the energies $E'_{\g_{+}}$ and $E'_{\g_{-}}$ are  no longer equal to each other because of the relativistic Doppler effect.}}}
\end{center}
\end{figure}

Einstein, in his first paper on special relativity \cite{Einstein1}, published a few months earlier, derived the relativisitic Doppler formula for light from the Lorentz transformations; in particular,   he established in that paper that if the radiation possesses  in $S$ a total energy $L$, then it possesses
a total energy $L'$ in $S'$ with $L$ and $L'$ related as follows.  
$$L' =  L \frac{1}{\sqrt{1-\frac{v^2}{c^2}}}. \eqno(10)$$
More generally, it follows, from Einstein's first special relativity paper \cite{Einstein1},  that if an electromagnetic pulse, i.e. plane electromagnetic wave, of energy $E(\varphi)$ is emitted by a particle at rest, then in an inertial reference frame where the particle moves with velocity $\vec v$ we have:   
$$E'(\varphi) = E(\varphi) \frac{1+\frac{v}{c}\cos(\varphi)}{\sqrt{1-\frac{v^2}{c^2}}}\eqno(11)$$
where $\varphi$ is the angle between the direction of the electromagnetic pulse and the velocity $\vec v$.

By making use of Eq. (10) and conservation of energy in the $S$ and $S'$ frames, respectively, Einstein obtains the following two equations:
$$ E_0 = E_1 + \halb L + \halb L , \eqno(12)$$
$$ H_0 = H_1 +  L \frac{1}{\sqrt{1-\frac{v^2}{c^2}}} .\eqno(13)$$
Next he argues that
$$ H_0 -E_0 = K_0 +C ,\eqno(14)$$
$$H_1 - E_1 = K_1 +C. \eqno(15)$$
where, according to Einstein, $C$ is an additive constant which ``does not change during the emission of light"  \cite{Einstein}.  In the Appendix we establish the following result, which we shall make use of later on. 
\\
\\
\textbf{Proposition}: \textit{The constant $C$ in Eqns. (14) and (15) is zero.}
\\
\\
From Eqns. (14) and (15) together with the previous two equations, which were obtained from conservation of energy, Einstein obtains
$$ K_0 - K_1 = L \left\{  \frac{1}{\sqrt{1-\frac{v^2}{c^2}}} -1 \right\} . \eqno(16)$$
This is his second to last equation.  His last equation is obtained from this equation by a Maclaurin series expansion of the right hand side neglecting terms of order higher than $v^2/c^2$.  It is
$$ K_0 - K_1 = \halb \frac{L}{c^2} v^2 . $$ 
To obtain Einstein's last statements, which describe mass energy equivalence, (i.e.the four sentences in Einstein's paper after ``From this equation it directly follows that:--") one simply uses the nonrelativistic formulae $K_0 = \halb m_0 v^2$ and $K_1 = \halb m_1 v^2$ where $m_1$ is the mass of the body after emission of radiation and $m_0$ is the mass of the body before the emission of radiation.  Substitution of these expressions for $K_0$ and $K_1$ into the above equation leads us to $$ m_0 - m_1 = \frac{L}{c^2} \eqno(17) $$
which Einstein did not bother to write down explicitly, but rather just described  it in words, exactly as did Poincar\'e in 1900.

\section{Further Development of Poincar\'e's Approach}
In this section we develop further Poincar\'e's treatment of his recoil process, which we described in Section 1, along lines similar to Einstein in Ref. \cite{Einstein}.
Poincar\'e's recoil process can viewed of as a decay process in which a particle, by emitting an  electromagnetic plane wave pulse  in a particular direction, decays into a recoiling particle with a slightly different mass. It is interesting that, by following Einstein,  we almost succeed in giving a completely relativistic treatment of the decay process.  Only at the very end of our analysis is it necessary, just as it is for Einstein, to make use of a non-relativistic approximation. 

Like Einstein,  we consider the process in different inertial frames, and we attempt, just as Einstein, to make use \textit{only} of energy conservation and Eqns. (14) and (15) in order to arrive at Poincar\'e's result on mass-energy equivalence, which, in equation form, is given by Eq. (8).   We are not successful in this attempt.  Just as we shall also show in Section 5 for Einstein's process, it is not possible to get at mass-energy equivalence solely out of arguments based on energy considerations.  We also must make use of  Poincar\'e's generalization of momentum conservation, Eq. (1), in order obtain $E_\g= m c^2$.

Fig. 1 shows Poincar\'e's  recoil/emission process in the rest frame $S$ of the  body, i.e. the Hertzian exciter,  before the emission of radiation. The mass of the body before and after emission of radiation are $M$ and $M_0$ respectively.  Its recoil speed is $v$. The mass of the radiation pulse is $m$.  Such a process occurs frequently in nuclear physics.  An examples is the radioactive decay of $^{57}$Co into $^{57}$Fe  plus a 14.4 KeV gamma ray, a decay which is commonly used in the M\"ossbauer effect. 

We also consider the process in a reference frame $S'$ where the body is seen to move with speed $v$ to the right before the radiation emission occurs.  This  is shown in Fig. 4. Since the recoil speed in $S$ is $v$, it follows that after the emission of radiation the body is at rest in $S'$.   

The  mass of the radiation as viewed by an observer in $S'$ is $m'$.   Combining Eq. (8)  for the mass of Poincar\'e's electromagnetic fluid, i.e. the mass of the radiation pulse in the rest frame of $M$, with Einstein's relativistic Dopper shift formula, i.e. Eq. (11), for the energies $E_\g$ and $E'_\g$ of a radiation pulse as measured in the $S$ and $S'$ frames, respectively, we get
$$ m' =  \frac{m(1+\frac{v}{c})}{\sqrt{1-\frac{v^2}{c^2}}} .\eqno(18)$$

\begin{figure}[htb]
\begin{center}
\begin{tikzpicture}[
    media/.style={font={\footnotesize\sffamily}},
    wave/.style={decorate,decoration={snake,post length=1.4mm,amplitude=.6mm,
        segment length=2mm},thick},
    interface/.style={postaction={draw,decorate,decoration={border,angle=-45,
                    amplitude=0.3cm,segment length=2mm}}},
   scale=1]

 \shade[shading=sphere] (-1,0) circle [radius=.5cm];

 \shade[shading=sphere] (0,2.5) circle [radius=.5cm];

 \shade[shading=sphere] (1,0) circle [radius=.3cm];

\node at (4,0) {$~~~~~~~~~~~~~~~~~$};

\node at (.7,2.5) [single arrow,draw,inner sep=.3,minimum width=.35cm,minimum height=.5cm,single arrow head extend=0.1cm] {};

\node at (-4.5, 3.5) {$S'$};

\node at (1.4,.6) {$E'_\gamma$};
\node at (-1.6,.6) {$E'_0$};

\node at (2.0,-.8) {$m'c$};
\node at (1.3,3.1) {$Mv$};

\node at (0, 3.5) {$\textrm{Before}$};

\node at (0, 1) {$\textrm{After}$};

\node at (1.8,0) [ single arrow, draw,inner sep=.3,minimum width=.35cm,minimum height=1cm,single arrow head extend=0.1cm] {};

\node at (-1, 2.9) {$E'$};

\node at (0, 2.5) {$M$};

\node at (1, 0) {$m$};
\node at (-1, 0) {$M_0$};

\end{tikzpicture}
\caption{Poincar\'e's recoil process as viewed in the inertial frame $S'$ moving with speed $v$ relative to the rest frame of the body before the emission of the light pulse occured. In $S'$ the body is seen initially moving to the right with speed $v$ where $v$ is the recoil speed of the body in Fig. 1. After the emission as occured the body appears to be at rest in $S'$.}
\end{center}
\end{figure}
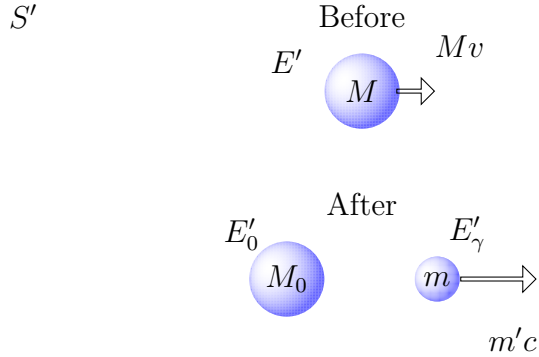
  
\noindent
Conservation of energy in $S$  gives 
$$ E= E_0+E_\gamma .$$
Conservation of energy in $S'$ gives 
$$ E'=E_0' + E_\g' $$

Eqns. (14) and (15) (Einstein's assumptions)  are general statements regarding energy and kinetic energy and they are applicable both non-relativistically and relativistically to any emission process, so that they can be applied to the situation described here. Furthermore, there is no problem here with ambiguity of interpretation of $K_0$, since, following Poincar\'e in \cite{Poincare}, we  assume conservation of momentum. Thus we have from Eqns. (14) and (15), with $C=0$ by the Proposition, that
$$ E' = E +K $$
and 
$$ E_0 = E_0' + K_0 $$
where $K$ and $K_0$ are, respectively, the relativistic kinetic energies of the masses $M$ and $M_0$ when they move with speed $v$.

Collecting the above results we obtain

\setcounter{equation}{18}

\begin{eqnarray} E= E_0+E_\gamma \\  E'=E_0' + E_\g'  \\ E' = E +K  \\ E_0 = E_0' + K_0  \end{eqnarray}
Substitution of Eqns. (20) and (21) into Eq. (19) 
gives
$$ E_0' + E_\g' = E_0+E_\gamma +K .$$
Using  Eq. (22) we replace $E_0$ by $E_0' + K_0 $ in this equation and simplify to obtain
$$ K+K_0 = E_\g'  - E_\g . \eqno(23)$$

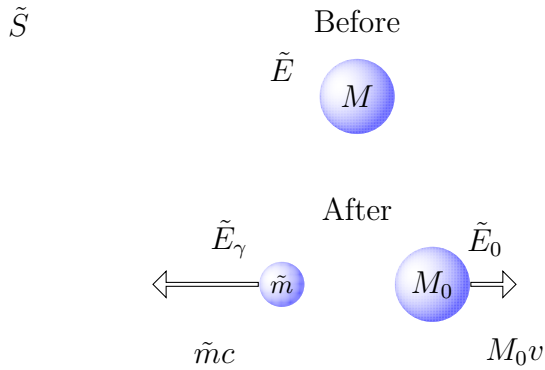
\begin{figure}[htb]
\begin{center}
\begin{tikzpicture}[
    media/.style={font={\footnotesize\sffamily}},
    wave/.style={decorate,decoration={snake,post length=1.4mm,amplitude=.6mm,
        segment length=2mm},thick},
    interface/.style={postaction={draw,decorate,decoration={border,angle=-45,
                    amplitude=0.3cm,segment length=2mm}}},scale=1]

 \shade[shading=sphere] (-1,0) circle [radius=.3cm];

\node at (4,0) {$~~~~~~~~~~~~~~~~~$};

 \shade[shading=sphere] (0,2.5) circle [radius=.5cm];

 \shade[shading=sphere] (1,0) circle [radius=.5cm];

\node at (-4.5, 3.5) {$\tilde S$};

\node at (1.7,.6) {$\tilde E_0$};
\node at (-1.7,.6) {$\tilde E_\gamma$};

\node at (-1.9,-.9) {$\tilde m c$};

\node at (2.1,-.9) {$M_0 v$};

\node at (0, 3.5) {$\textrm{Before}$};

\node at (0, 1) {$\textrm{After}$};

\node at (1.8,0) [ single arrow, draw,inner sep=.3,minimum width=.35cm,minimum height=.6cm,single arrow head extend=0.1cm] {};

\node at (-2.0,0) [single arrow,draw,inner sep=.3,minimum width=.35cm,minimum height=1.4cm,single arrow head extend=0.1cm, rotate=180] {};

\node at (-1, 2.9) {$\tilde E$};

\node at (0, 2.5) {$M$};

\node at (1, 0) {$M_0$};

\node at (-1, 0) {{\footnotesize${\tilde m}$}};

\end{tikzpicture}
\caption{Mirror image of Fig. 1.  We put tildes over quantities such energies, even though by reflection symmetry they must necessarily be the same as in Fig. 1.}
\end{center}
\end{figure}

To make further progress, we need to consider the mirror image of Fig. 1 which is Fig. 5.  This  can be interpreted  as the the process shown in Fig. 1 but as seen in a reference frame $\tilde S$ connected to $S$ by a reflection about the vertical symmetry axis i.e. about a vertical line in Fig. 1 which passes through $M$.  Conservation of energy in $\tilde S$  gives 
$$\tilde E= \tilde E_0+\tilde E_\gamma . $$
Finally, we view the process  from a reference frame $\tilde S'$ moving with velocity $-\vec v$ with respect to $\tilde S$, so that the body, which is at rest in $\tilde S$,  is seen in $\tilde S'$ to move with speed $v$ to the right before the radiation emission occurs.  This is shown in Fig. 6.  Since the recoil speed in $\tilde S$ is $v$, it follows that after the emission of radiation the body moves to the right with speed $\tilde v =\frac{2v}{1+\frac{v^2}{c^2}}$ in $\tilde S'$ according to the relativistic addition of velocities formula in special relativity.

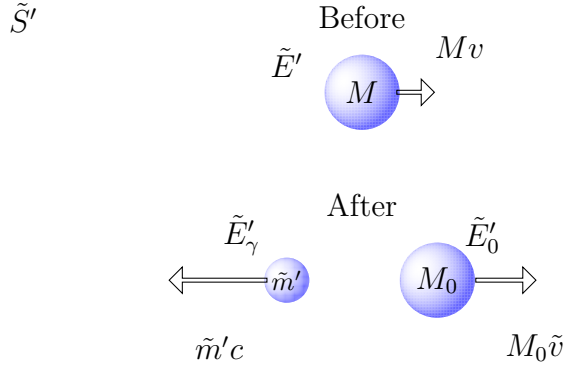
\begin{figure}[htb]
\begin{center}
\begin{tikzpicture}[
    media/.style={font={\footnotesize\sffamily}},
    wave/.style={decorate,decoration={snake,post length=1.4mm,amplitude=.6mm,
        segment length=2mm},thick},
    interface/.style={postaction={draw,decorate,decoration={border,angle=-45,
                    amplitude=0.3cm,segment length=2mm}}},scale=1]

 \shade[shading=sphere] (-1,0) circle [radius=.3cm];

 \shade[shading=sphere] (0,2.5) circle [radius=.5cm];

 \shade[shading=sphere] (1,0) circle [radius=.5cm];

\node at (4,0) {$~~~~~~~~~~~~~~~~~$};

\node at (.7,2.5) [single arrow,draw,inner sep=.3,minimum width=.35cm,minimum height=.5cm,single arrow head extend=0.1cm] {};

\node at (-1, 0) {\footnotesize{$\tilde m'$}};

\node at (-4.5, 3.5) {$\tilde S'$};

\node at (1.6,.6) {$\tilde E'_0$};
\node at (-1.6,.6) {$\tilde E'_\gamma$};

\node at (-1.9,-.9) {$\tilde m'c$};
\node at (1.3,3.1) {$Mv$};

\node at (2.3,-.9) {$M_0\tilde v$};

\node at (0, 3.5) {$\textrm{Before}$};

\node at (0, 1) {$\textrm{After}$};

\node at (1.9,0) [ single arrow, draw,inner sep=.3,minimum width=.35cm,minimum height=.8cm,single arrow head extend=0.1cm] {};

\node at (-1.9,0) [single arrow,draw,inner sep=.3,minimum width=.35cm,minimum height=1.3cm,single arrow head extend=0.1cm, rotate=180] {};

\node at (-1, 2.9) {$\tilde E'$};

\node at (0, 2.5) {$M$};

\node at (1, 0) {$M_0$};

\end{tikzpicture}
\caption{Mirror image of Fig. 1  viewed from an inertial frame ${\tilde S}'$ moving with speed $v$ relative to $\tilde S$.  The body, which is at rest in $\tilde S$,  is seen in $\tilde S'$ to move with speed $v$ to the right before the radiation emission occurs. }

\end{center}
\end{figure}

 Conservation of energy in $\tilde S'$ gives 
$$ \tilde E' = \tilde E_0' + \tilde E_\gamma' .$$
For the mirror image situation, the mass $M_0$ is neither at rest in  $\tilde S$ nor in $\tilde S'$, so that we get only one Einstein condition: 
$$ \tilde E' = \tilde E + \tilde K  .\eqno(24)$$  
From the figures it is clear that $\tilde E_0 = E_0$, $\tilde E = E$ ($\Rightarrow \; E_\g = \tilde E_\g$ by conservation of energy, and thus $\tilde m = m$) and $\tilde E' = E'$.  Thus $\tilde K =K$ and Eq. (24) is the same as Eq. (21).  However, comparing Figs. (4) and (6) we see that, according to Einstein:
$$ \tilde E_0'  = E_0' + \tilde K_0 $$
where $\tilde K_0$ is the kinetic energy of $M_0$ when it moves with speed $\frac{2v}{1+\frac{v^2}{c^2}} \approx 2v$.

\setcounter{equation}{24}
Collecting these results for the tilde analysis leads us to the following set of equations
\begin{eqnarray} E= \tilde E_0+E_\gamma = E_0 +E_\g  \quad \\E' = \tilde E' =  \tilde E_0' + \tilde E_\gamma'  \qquad  \\  E'=  \tilde E' = \tilde E + \tilde K  = E + K  \\ \tilde E_0'  = E_0' + \tilde K_0  \qquad . \end{eqnarray}
 
Subtracting Eq. (25) from Eq. (26) gives
$$ K=E'-E=\tilde E_0' +\tilde E_\g' -E_0 - E_\g .$$
Now use Eq. (22) to replace $E_0$ in this equation by $E_0'+K_0$ to obtain
$$K=E'-E=\tilde E_0' +\tilde E_\g' -E_0'-K_0 - E_\g = \tilde K_0 -K_0 +\tilde E_\g'  - E_\g$$
where in obtaining the last term we used Eq. (28).
Thus 
$$ K + K_0 = \tilde K_0 + \tilde E_\g'  - E_\g. $$
With the help of  Eq. (23) we finally obtain
$$E'_\g - E_\g  = \tilde K_0 + \tilde E'_\g  - E_\g$$
which simplifies to
$$ \tilde K_0 = E'_\g - \tilde E'_\g   .\eqno(29)$$
Now $E_\g'$ is the energy of a light pulse from a source moving with speed $v$ in the same direction as the light pulse and $\tilde E'_\g$ is the energy of a light pulse emitted from a source moving with speed $v$ in the opposite direction as the light pulse.  Hence, according to Eq. (11),
$$ \tilde K_0 = \frac{E_\g (1+\frac{v}c)}{\sqrt{1- \frac{v^2}{c^2}}} - \frac{E_\g (1-\frac{v}c)}{\sqrt{1- \frac{v^2}{c^2}}} = \frac{2 E_\g\frac{v}c}{\sqrt{1- \frac{v^2}{c^2}}}. \eqno(30) $$

Up to this point everything has been relativistically exact.   Approximating Eq. (30) nonrelativistically, we obtain $E_\g \frac{2v}c$, to lowest order in $v/c$, for the right hand side of the equation. 
We equate this to the nonrelativistic kinetic energy of a particle of mass $M_0$ moving with speed $2v$, from which we  obtain\footnote{Field in his paper makes use of a definition of the rest  mass of a body in terms of its relativistic kinetic energy, which he attributes to Stachel and Torretti \cite{Stachel}.    In Ref. \cite{Stachel}, Stachel and Torretti define the mass of an object to be  
$$ M= \lim\limits_{v \rightarrow 0} \frac{K}{v^2/2} \eqno(31) $$
where $M$ is the mass of the particle and $K$ is its relativistic kinetic energy when it is moving with constant speed $v$.  They do not place any limitations on applicability of Eq. (31) and they claim it serves as a general definition for the inertial mass of an object when we are given the relativistic kinetic energy of the object. 

Unfortunately, there seems to be a problem with their definition when applied to this example.  If we apply it to Eq. (30), remembering that for  the kinetic energy $\tilde K_0$ the speed of $M_0$ is $\tilde v =2v$, we obtain
$$M_0 = \lim\limits_{\tilde v \rightarrow 0} \frac{\tilde K_0}{{{\tilde v}^2/2}}  = \lim\limits_{\tilde v \rightarrow 0}\frac{\frac{2 E_\g\frac{v}c}{\sqrt{1- \frac{v^2}{c^2}}}} {{\tilde v^2/2}} = \lim\limits_{v \rightarrow 0}\frac{\frac{2 E_\g\frac{v}c}{\sqrt{1- \frac{v^2}{c^2}}}} {{ 2v^2}} = \lim\limits_{v \rightarrow 0} \frac{E_\g}{vc    \sqrt{1- \frac{v^2}{c^2}}} = \infty $$
assuming $E_\g \ne 0$. ($E_\g=0$ means no decay occured, since conservation of energy with $E_\g =0$ imples $E =E_0$ according to Eq. (19).)
This is clearly a nonsensical result.  So we conclude that Stachel and Torretti's definition of inertial mass does not have general validity. 

A better definition of the inertial mass of an object might perhaps involve following Wigner and In\"on\"u \cite{Wigner}, that is to say, letting $c\rightarrow \infty$ simultaneously as the energies of the particles go to infinity in such a way that their ratios with $c$ remain finite.  For the case at hand this would amount to something along the following lines.  Instead of letting $v\rightarrow 0$, let $E_\g \rightarrow \infty$, $E_0 \rightarrow \infty$ and $c \rightarrow \infty$ in such a way that $\frac{E_\g}{c} \rightarrow p_\g$ where $p_\g = mc$. Then 
 
$$M_0 = \lim\limits_{\begin{array}c c \rightarrow \infty \\ E_\g \rightarrow \infty \\E_0 \rightarrow \infty \\ \frac{E_\g}{c}\rightarrow p_\g \nonumber \end{array} } \frac{\tilde K_0}{{{\tilde v}^2/2}} =  \lim\limits_{\begin{array}c c \rightarrow \infty \\ E_\g \rightarrow \infty \\E_0 \rightarrow \infty \\ \frac{E_\g}{c}\rightarrow p_\g \nonumber \end{array} } \frac{\frac{E_\g}c}{v    \sqrt{1- \frac{v^2}{c^2}}} = \frac{p_\g}v = \frac{mc}{v} .$$
With $m$ given by Eq. (8) this will then agree with Eq. (32).   In any case, it seems best to avoid using Stachel and Torretti's definition of mass, i.e. Eq. (31).}

$$ M_0 = \frac{E_\g}{vc} . \eqno(32)$$
This seem to be as far as arguments,  based solely on energy considerations, can take us.  Eq. (32) confirms only Poincar\'e's result  that  the nonrelativistic momentum $M_0v$ of the recoiling particle is equal to the energy $E_\g$ of the radiation pulse divided by $c$. It does not seem possible to obtain Poincar\'e's result on mass energy equivalence for radiation, i.e. $E_\g = mc^2,$  using only energy considerations.

The additional input necessary in order to obtain Eq. (8) is provided by Eq. (1), Poincar\'e's generalization of the law of conservation of momentum.  It led us to  
Eq. (9) from which we obtain
$$ M_0 v = mc \eqno(33)$$
giving together with Eq. (32) 
$$ E_\g = mc^2. \eqno(34)$$

It could have been possible to think that kinetic energy along with energy considerations would have avoided the use of momentum considerations, since it is a derived concept obtained from momentum and velocity.\footnote{Kinetic energy, or \textit{vis viva} as its originator, Leibnitz, called it, is up to a numerical factor the product of the momentum times the velocity of an object.  The way it is introduced in some of today's college physics textbooks and the way Stachel and Torretti \cite{Stachel} and also Field \cite{Field} define it, as the work done by the net force on the object, came much later around the mid-19th century \cite{Jammer}, \cite{Jammer2}.}     For example, the nonrelativistic kinetic energy of $M_0$ in the rest frame of $M$ is $$ K_0 = \halb M_0 v^2 = \halb |\vec v \cdot \vec p| \eqno(35) $$
and the expression for it in any other frame can be obtained from this equation by applying a  Galilean transformation to $\vec v$ and $\vec p$.  Specifically, under a Galilean boost to the frame $S'$, we obtain
$$ K_0' = \halb |\vec v\, ' \cdot \vec p'| =\halb |\vec v\,' \cdot m {\vec v}\,'| =  \halb |(\vec v -\vec v ) \cdot m(\vec v -  \vec v)| =0 . \eqno(36) $$ 
Eq. (35) shows that we can always replace considerations about kinetic energy by considerations regarding momentum and velocity.  However, the converse statement is weaker, since  momentum also carries directional information involved in momentum balance.   In particular, only with Poincar\'e's generalization of the law of momentum conservation, Eq. (1),  which includes in addition to the momenta of ponderable matter a term involving radiation, can we interpret $\frac{E_\g}{c}$ as radiation momentum given by Eq. (9), i.e. $\frac{E_\g}{c} = p_\g = mc$ where $m$ is the mass of radiation.

Poincar\'e's approach to mass-energy equivalence  relies on momentum conservation and momentum of radiation in obtaining $E_\g=mc^2$. Our analysis in this section shows  that  it is not possible to get at any clear statement regarding mass-energy equivalence by considering decay processes, such as the one considered by Poincar\'e, by making use only of energy conservation and energy considerations.  Momentum considerations also play an indispensible role.

 \section{On Einstein's Assumption of the Equality of the Energies of Two Light Pulses in the Rest Frame of the Particle}

The process of emission of radiation by a stationary body, described by Einstein in his paper, must have seemed shocking to a Maxwellian physicist at the beginning of the twentieth century.  They  would certainly have doubted such a process as the one Einstein proposed to be possible, for, as Larmor and others had just convinced them,  a stationary body cannot give off electromagnetic radiation.  

On the other hand, the assumption that the energies of the two light pulses in the rest frame $S$ of the particle are the same, seems to have, for the most part, gone unoticed.   However,  
assuming the equality of the two energies, without providing sufficient justification as to why, is at odds with the relativity of motion principle.  
To declare by \textit{fiat} that the electromagnetic energy from emission of light along a particular direction is the same as in the opposite  direction as viewed by an observer in the rest frame of the body singles out a privileged frame of reference, namely, the rest frame of the body. \textit{Why should this reference frame be distinguished from  all the others by only it having the right to declare the two energies equal to one another?}  
We are given no information as to the internal constitution of the body before or after the emission, nor does Einstein give any information about the two light pulses other than their energies. Momentum considerations are absent; in particular, there is nothing about the momentum of radiation and its inertia. Under such circumstances, the relativity principle demands that we should consider it equally possible to equate the two energies of the two light pulses in the $S'$ frame.  Just as in the twin paradox, where either twin has the right to assert that he is stationary and it is the other twin moving with constant velocity relative to him who didn't age, so also here, both scenerios regarding equality of the two energies should be considered as equally likely possibilities in the emission process considered by Einstein.  An observer in $S'$ has just as much right to insist that he is the one who measures the two energies as equal, just as an observer in $S$ can claim. 

So, if we are to make use of nothing more than conservation of energy, then the relativity principle demands that we also consider the situation where we have equality of the two energies $E'_{\g_+}$ and  $E'_{\g_-}$, the energies of the two photons in the reference frame $S'$ of Fig. 3. i.e. $E'_{\g_+} =  E'_{\g_-}$.    In other words, $E_{\g_+} =  E_{\g_-}$ in  $S$ or $E'_{\g_+} =  E'_{\g_-}$ in $S'$, should be considered equally possible.  Let us repeat Einstein's analysis for the situation $E'_{\g_+} =  E'_{\g_-}$ in $S'$ and see what it leads us to.  
Instead of Eq. (1) we now have 
$$ L =  L' \frac{1}{\sqrt{1-\frac{v^2}{c^2}}}. \eqno(10^\textrm{bis})$$
Conservation of energy in $S$ and in $S'$, respectively, gives

$$ E_0 = E_1  + E_{\g_+} +  E_{\g_-} = E_1 +L' \frac{1}{{\sqrt{1-\frac{v^2}{c^2}}}}, \eqno(12^\textrm{bis})$$
and
$$ H_0 = H_1 +   E'_{\g_+} +  E'_{\g_-} = H_1 + L'\eqno(13^\textrm{bis})$$
since, now we have $E'_{\g_+} =  E'_{\g_-} = \halb L'$.
It is clear that Eqns. (14) and (15), remain the same, since they are simply assumptions relating the kinetic energy and total energies of the body in the two reference frames before and after the emission process.  Thus Eqns. (12$^\textrm{bis}$),  (13$^\textrm{bis}$), (14) and (15) lead us to
 $$ K_0 - K_1 = L \left\{{\sqrt{1-\frac{v^2}{c^2}}} -1 \right\} . \eqno(16^\textrm{bis})  $$
Repeating the final steps in Einstein's analysis described at the end of section 3 leads us to the equation
$$ m_0-m_1 = -\frac{L}{c^2}. \eqno(17^\textrm{bis}) $$
Since $L$ is a positive number this equation implies that $m_1 > m_0$, which would imply that the inertial mass of a body increases when it gives off radiation provided $m_0$ and $m_1$ are both positive.  On the other hand, if we allowed for negative inertial masses, as some have entertained to be true of antiparticles  \cite{Debergh}, then, with both $m_1$ and $m_0$  negative numbers, we would have  that $|m_1| < |m_0|$.

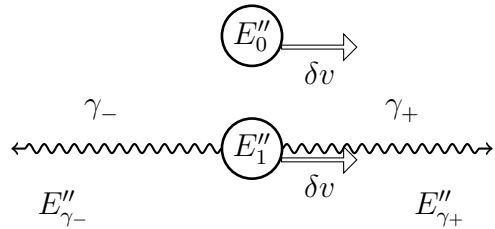
\begin{figure}[htb]
\begin{center}
\begin{tikzpicture}[
    media/.style={font={\footnotesize\sffamily}},
    wave/.style={
        decorate,decoration={snake,post length=1.4mm,amplitude=.6mm,
        segment length=2mm},thick},
    interface/.style={
        postaction={draw,decorate,decoration={border,angle=-45,
                    amplitude=0.3cm,segment length=2mm}}},
    ]
]

    \draw[->,wave]  
(-180:.4cm)--(-180:3.2cm)node[left]{$~$};

\node at (.88,-.15) [single arrow,draw,inner sep=.3,minimum width=.35cm,minimum height=1cm,single arrow head extend=0.1cm] {};
\node at (.88,1.35) [single arrow,draw,inner sep=.3,minimum width=.35cm,minimum height=1cm,single arrow head extend=0.1cm] {};

\node at (-4.5, 3) {$S''$};

    \draw[line width=1pt](0,1.5)circle(.4cm);
\node at (.9, -.5) {$ \d v$};
  
    \draw[->,wave]
       (0:.4cm)--(0:3.2cm)node[right]{};

\node at (-2,+.5) {$\gamma_-^{}$};

\node at (.9, 1.0) {$\d v$};
\node at (0, 1.5) {$E''_0$};

\node at (0, 0) {$E''_1$};

\node at (2.5,-.8) {$E''_{\gamma_+}$};
\node at (-2.5,-.8) {$E''_{\gamma_-}$};
'
\node at (2,+.5) {$\gamma_+^{}$};

    \draw[line width=1pt](0,0)circle(.4cm);
    
\end{tikzpicture}
\caption{\footnotesize{Emission of two plane light pulses from a body as viewed in the inertial frame $S''$ which moves with arbitrary speed $\d v$ relative to the rest frame $S$ of the body, where $\d$ can be any positive number such that $\d v < c$ and $v = |\vec v |$ with $\vec v$ the velocity of the body as viewed by an observer in $S'$. The light pulse moving to the right has energy $E''_{\g_{+}}$ and the other one has energy $E''_{\g_{-}}$. In $S''$ the total energy of the body before the emission of light is $E''_0$ and the energy of the body after the emission of the two light pulses is $E''_1$.  
}}
\end{center}
\end{figure}

In fact, the only way to do complete justice to the relativity principle is to demand that the inertial frame where we set the two energies equal to one another should  \textit{ab initio} be completely arbitrary. So we should consider the emission of the two plane light pulses from the body as viewed in the inertial frame $S''$ depicted in Fig. 7 moving with arbitrary velocity $-\d \vec v$ relative to the rest frame $S$ of the body.  An observer in  $S''$ then sees the body move to the right with speed $\delta v$ both before and after the emission process.  $\d$ is a real number subject to the following limitations.  $\d \vec v$ could be any intermediate velocity between zero and $\vec v$ or we could even consider negative velocities and velocities greater than $\vec v$. However, if $v$ is viewed as arbitrary subject only to $v< c$, then this implies that $\d$ should satisfy $0\le \d\le 1$, in order that $\d v$ for any $v$ is never larger than $c$.  

According to the Doppler formula, Eq. (11), the energies of the two photons as measured by an observer in $S''$ in terms of the rest frame energies $E_{\g_+}$ and $E_{\g_-}$ are
$$  E''_{\g_+} = \frac{E_{\g_+}\left(1+\frac{\d v}{c}\right)}{\sqrt{1- \frac{(\d v)^2}{c^2}}}  \quad, \quad E''_{\g_-} = \frac{E_{\g_-}\left(1-\frac{\d v}{c}\right)}{\sqrt{1- \frac{(\d v)^2}{c^2}}} \eqno(36) $$
and the inverse relations are

$$  E_{\g_+} = \frac{E''_{\g_+}\left(1-\frac{\d v}{c}\right)}{\sqrt{1- \frac{(\d v)^2}{c^2}}}  \quad, \quad E_{\g_-} = \frac{E''_{\g_-}\left(1+\frac{\d v}{c}\right)}{\sqrt{1- \frac{(\d v)^2}{c^2}}} .  \eqno(37)$$
Likewise, the energies $E'_{\g_+} $ and $E'_{\g_-} $ of the two wave pulses in the $S'$ frame depicted in Fig. 2 are
$$  E'_{\g_+} = \frac{E_{\g_+}\left(1+\frac{v}{c}\right)}{\sqrt{1- \frac{v^2}{c^2}}}  \quad, \quad E'_{\g_-} = \frac{E_{\g_-}\left(1-\frac{v}{c}\right)}{\sqrt{1- \frac{v^2}{c^2}}}  .\eqno(38)$$
From these equations we obtain

$$  E'_{\g_+} = \frac{E''_{\g_+}\left(1+\frac{v}{c}\right)\left(1-\frac{\d v}{c}\right)}{\sqrt{1- \frac{v^2}{c^2}}\sqrt{1- \frac{(\d v)^2}{c^2}}}  \eqno(39)$$
and
$$E'_{\g_-} = \frac{E''_{\g_-}\left(1-\frac{v}{c}\right)\left(1+\frac{\d v}{c}\right)}{\sqrt{1- \frac{v^2}{c^2}}\sqrt{1- \frac{(\d v)^2}{c^2}}}  .\eqno(40)$$ 
Equating the energies of the two photons in the $S''$ frame yields 
$$ E''_{\g_+} = E''_{\g_-} =: \halb L'' \eqno(41)$$
This together with Eqns. (39) and (40) gives
$$  E'_{\g_+} + E'_{\g_-} = \frac{E''_{\g_+}\left(1+\frac{v}{c}\right)\left(1-\frac{\d v}{c}\right)}{\sqrt{1- \frac{v^2}{c^2}}\sqrt{1- \frac{(\d v)^2}{c^2}}}   + \frac{E''_{\g_-}\left(1-\frac{v}{c}\right)\left(1+\frac{\d v}{c}\right)}{\sqrt{1- \frac{v^2}{c^2}}\sqrt{1- \frac{(\d v)^2}{c^2}}} =$$
$$ = \halb L'' \left\{ \frac{  \left(1+\frac{v}{c}\right)\left(1-\frac{\d v}{c}\right) + \left(1-\frac{v}{c}\right)\left(1+\frac{\d v}{c}\right)}{\sqrt{1- \frac{v^2}{c^2}}\sqrt{1- \frac{(\d v)^2}{c^2}}}\right\} = L''\frac{\left(1-\frac{\d v^2}{c^2}\right)}{\sqrt{1- \frac{v^2}{c^2}}\sqrt{1- \frac{(\d v)^2}{c^2}}}  \eqno(42)$$
From Eqns. (37)  and (41) we obtain
$$ L =  E_{\g_+} + E_{\g_-} = \frac{L''}{ \sqrt{1- \frac{(\d v)^2}{c^2}}} $$
and thus
$$ L'' = \sqrt{1- \frac{(\d v)^2}{c^2}} L .  \eqno(43)$$
Conservation of energy in $S$ and $S'$ and Eqns. (42) and (43) leads us to
$$ E_0 = E_1 + E_{\g_+} + E_{\g_-} = E_1 + L \eqno(12^\textrm{ter})$$
and
$$ H_0 = H_1 + E'_{\g_+} + E'_{\g_-} = H_1 + L \frac{ \left(1-\frac{\d v^2}{c^2}\right)}{\sqrt{1- \frac{v^2}{c^2}}}, \eqno(13^\textrm{ter}).$$
(As in the other two cases, it is the energies in the $S$ and $S'$ frames that we are comparing, not the energies in the $S$ and $S''$ frames.  The only role $S''$ plays in the analysis is its use as an arbitrary reference frame in which the two photon energies are set equal to one another.)  Finally, using  Eqns. (12$^\textrm{ter}$), (13$^\textrm{ter}$), (14) and (15) we obtain 
$$ K_0 - K_1 = L\left\{ \frac{ 1- \frac{\d v^2}{c^2}}{\sqrt{1- \frac{v^2}{c^2}}} - 1\right\} .\eqno(16^\textrm{ter})$$
Repeating Einstein's nonrelativistic approximations we obtain
$$ m_0 - m_1 = (1-2\d) \frac{L}{c^2} . \eqno(17^\textrm{ter})$$
Since $\d$ is arbitrary this does not lead us to any definite and unique relationship between the inertia of a body and its energy content. We can obtain any value we wish for the mass difference $m_0 - m_1$ depending upon the choice of $\d$.  For $\d=0$ we obtain Eq. (17), i.e. $E= mc^2$ with $E= L$ and $m=m_0-m_1$.  For $\d=1$ we obtain Eq. (17$^\textrm{bis}$).  By an appropriate choice of $\d$ ($\d= \frac{1}{8}$), we can even obtain $E= \frac{4}{3} m c^2$, i.e. the factor of 4/3 occuring in the electromagnetic mass which is also the Hasen\"ohrl  result \cite{Hasenoehrl}, \cite{Hasenoehrl1} on mass-energy equivalence.\footnote{Fritz Hasen\"ohrl, the Ph.D. adviser of Schr\"odinger at the University of Vienna,  published his papers on mass-energy equivalence in the \textit{Annalen der Physik} in the year prior to Einstein's 1905 article.  His paper, On the Theory of Radiation in Moving Bodies \cite{Hasenoehrl},  provides a much more complete analysis of mass-energy equivalence than Einstein's, making use of both energy and momentum considerations involving radiation momentum and pressure.  He considers a cavity containing radiation energy, and, like Einstein, he views it from two reference frames, one when the cavity is stationary and the other when it is moving with constant velocity.  In his paper, Hasen\"ohrl answers in the affirmative the question posed by Einstein in the title of his paper.  Boughn and Rothman  \cite{Rothman} seem to be in agreement with Ives \cite{Ives} that Einstein was most likely aware of Hasen\"ohrl's articles at the time he wrote his first paper on $E=mc^2$.  

Due to his approach Hasen\"ohrl was obliged to follow the theories of Abraham \cite{Abraham1}, \cite{Abraham2} and Lorentz \cite{Lorentz} regarding electromagnetic momentum and energy, and thus he obtained their result containing the factor of 4/3 for the electromagnetic mass, instead of $E=mc^2$.  The explanation for the discrepancy would have to wait until Poincar\'e \cite{Poincare2} in 1906 and then Fermi \cite{Fermi} in 1922 cleared up the matter: Poincar\'e, by conceiving of an extensible model of the electron containing an additional  field,  non-electromagnetic in origin or otherwise, exerting a constant negative pressure (Poincar\'e pressure) in the interior of the electron and vanishing outside of it \cite{Janssen}, and thus leading to a stabilizing attractive force only at the electron's surface, akin to the Casimir force \cite{Casimir} on a spherical conductor, except that the Casimir force is repulsive instead of attractive \cite{Boyer}; and Fermi, by defining the energy and the momentum in a relativistically covariant way \cite{Rohrlich}, \cite{Moylan}.  Griffiths in Ref. \cite{Griffiths} describes the Fermi approach as apparently still having ``a residue of confusion" about it.  In Ref. \cite{Boughn} Boughn tries to make Hasenr\"ohrl's analysis relativistically correct by incorporating Fermi's covariant approach into Hasen\"ohrl's treatment. }

 If we insist on setting the energies of the two wave pulses equal to one another in a particular frame, then, in order to uphold the relativity principle, we must insist that we can do that same procedure in any other inertial frame, and any one of the above possibilities results out of a correct treatment of Einstein's analysis.  Thus, \textit{it is not possible to get at $E=mc^2$ by considering emission of light from a body using arguments based solely on conservation of energy and the relativity of motion principle}. The relativity principle demands that all inertial reference frames are treated on a equal footing, and, as we have shown, our insistence on upholding this principle leads us to other possibilities with which we have to contend, provided we only make use of energy conservation and Einstein's relations relating energy to kinetic energy.

We need some additional physical input to lead us to the right choice for $\d$ i.e. $\d=0$.  The missing ingredients are, of course, conservation of momentum together with Poincar\'e's momentum of radiation.  The analysis of the previous section, which involves the decay of a massive particle into a light pulse and another particle,  was meant to help make clear why this is so.  Surely, for the Einstein example considered in this section, no one will argue with the fact that equating  the energies of the two light pulses in the rest frame of the decaying particle involves implict use of momentum conservation,  provided one accepts the fact that radiation carries off momentum.   Futhermore, \textit{conservation of momentum in an arbitrary reference frame justifies Einstein's fiat declaration that $E_{\g_+} =  E_{\g_-}$}.  To prove these claims we argue as follows.  Using the approximate nonrelativistic formula for the momentum of the body together with Poincar\'e's momentum of radiation, conservation of momentum in the $S''$ frame gives, according to Fig. 7, with $E''_{\g_+} =  E''_{\g_-}= \halb L''$:
$$ m_0 \d v = m_1 \d v +  \frac{E''_{\g_+}}{c} -  \frac{E''_{\g_-}}{c} = m_1 \d v +\halb \frac{L''}c - \halb \frac{L''}c = m_1 \d v.$$
For $v \ne 0$, this equation implies either $m_0=m_1$, which means no emission of radiation whatsoever, and hence nothing happens, or  $\d=0$, which implies by Eqns. (36) and (37) that $E_{\g_+} =  E_{\g_-}$, which is Einstein' assumption of the equality of the energies of the two light pulses in the rest frame of the body.  The case $v=0$ it is the trivial situation where $E_{\g_+} = E'_{\g_+}= E''_{\g_+}= E''_{\g_-} =  E'_{\g_-}= E_{\g_-}$ by Eqns. (36), (37) and (38), and hence again  
$E_{\g_+} =  E_{\g_-}$.\footnote{It follows from the Galilean transformations that conservation of momentum in one reference frame implies that conservation momentum should hold true in any other inertial reference frame, nonrelativistically, and, thus,  also hold true to  lowest nontrivial order of approximation in $v/c$, for the process considered here.  However, in order to make it absolutely clear that this is so, let us show directly how momentum conservation  in $S'$  would again lead to Einstein's assumption that $E_{\g_+} =  E_{\g_-}$. 

Conservation of momentum to order $v/c$ in $S'$ (cf. Fig. 3) is 
$$m_0 v = m_1 v + \frac{E'_{\g_+}}c -  \frac{E'_{\g_-}}c . \eqno(44)$$
Thus. from Eqns. (39) and (40)
$$(m_0 -m_1) v  = \frac{E''_{\g_+}\left(1+\frac{v}{c}\right)\left(1-\frac{\d v}{c}\right)}{\sqrt{1- \frac{v^2}{c^2}}\sqrt{1- \frac{(\d v)^2}{c^2}}}   - \frac{E''_{\g_-}\left(1-\frac{v}{c}\right)\left(1+\frac{\d v}{c}\right)}{\sqrt{1- \frac{v^2}{c^2}}\sqrt{1- \frac{(\d v)^2}{c^2}}} =$$
$$ = \frac{L''}{2c} \left\{ \frac{\left(1+\frac{v}{c}\right)\left(1-\frac{\d v}{c}\right)}{\sqrt{1- \frac{v^2}{c^2}}\sqrt{1- \frac{(\d v)^2}{c^2}}}   - \frac{\left(1-\frac{v}{c}\right)\left(1+\frac{\d v}{c}\right)}{\sqrt{1- \frac{v^2}{c^2}}\sqrt{1- \frac{(\d v)^2}{c^2}}}  \right\} = \frac{L''v }{c^2} 
 \frac{(1- \d)}{\sqrt{1-\frac{v^2}{c^2}}} ,$$
since $E''_{\g_+} =  E''_{\g_-}= \halb L''$. 
To lowest order in $v/c$ this gives
$$(m_0 -m_1) v  = \frac{Lv(1-\d)}{c^2} = \frac{Lv(1-2\d)}{c^2}+ \frac{Lv\d}{c^2}$$
Using Eq. (17$^\textrm{ter})$ gives
$$(m_0 -m_1) v  = (m_0 -m_1) v  + \frac{Lv\d}{c^2} $$
which implies $\frac{Lv\d}{c^2}=0$ and hence either $v=0$ or $\d=0$. The first choice, i.e. $v=0$, leads by Eq. (44) to $E'_{\g_+} =  E'_{\g_-}$.  But since $v=0$, $E'_{\g_+} =  E_{\g_+}$ and $E'_{\g_-} =  E_{\g_-}$ and, hence, Einstein's assumption.  The second choice $\d=0$ leads us, by Eqns. (36) and (37),  again to $E_{\g_+} =  E_{\g_-}$ since $E''_{\g_+} =  E''_{\g_-}$ by assumption.}

The reason that now there is no longer any conflict with the relativity of motion principle is because we are assuming equality of the energies of the two light pulses in an \textit{arbitrary inertial frame} $S''$, moving with an arbitrary velocity relative to the rest frame of the body, and then application of the law of momentum conservation forces us to $\d =0$ and hence  to $E_{\g_+} =  E_{\g_-}$.

One could possibly interpret Einstein's text to imply that Einstein was not aware that Poincar\'e in \cite{Poincare} had endowed radiation with mass and momentum, for he writes: ``\textit{If a body gives off the energy $L$ in the form of radiation, its mass diminishes by $L/c^2$} [emphasis by Einstein].  The fact that the energy withdrawn from the body becomes energy of radiation evidently makes no difference, \dots ."   Einstein never made any mention whatsoever of the word ``momentum" or of momentum conservation in his paper.

If it is true that Einstein, at that time, did not think it necessary to endow radiation with momentum and did not understand the need for momentum conservation in his analysis, then on what grounds was he justified in setting the two energies equal to each other?   Those of us, who are content with Einstein equating the two energies in the rest frame $S$ without making use of momentum conservation, should provide some other equally convincing reason for assuming them equal.   If  they honestly can do this, then, as I will explain in the next section,  they would be justified in upholding Einstein's derivation as correct, and the Ives' criticism as unjustified.

\section{ The Ives Criticism Revisited}

In this section we want to try to explain the logic of the Ives criticism of Einstein's paper.   In view of several attempts by supposedly reputable people to discredit Ives' criticism, it is clear that we need to be as careful as possible in explaining how we judge it to be valid.  For this reason we consider the  simplest possible case where the Einstein analysis is applicable; it is the situation in which the particle ceases to exist after the emission of radiation.  For this case, $E_1$, $H_1$ and $K_1$ are all zero.   An example of such a process occurs in nuclear physics;  it is neutral pion decay:
$$\pi_0 \to \gamma + \gamma . $$
This is the main decay mode of the neutral pion with a branching ratio of .99823.  It is also mentioned by Field for the purpose of simplifying argumentation and also to suggest the far-reaching implications of Einstein's emission process.  The second most common decay mode is $\pi_0 \rightarrow \g + e^+ + e^-$.  Even rarer decay modes are $\pi^0  \rightarrow e^+ +  e^-$ and $\pi^0 \rightarrow e^+ + e^- +  e^+  + e^-$.

Failure to take into account momentum considerations in emission processes such as Einstein's can lead to some pretty ridiculous conclusions.    In principle, Einstein's analysis applies to any form of wave or particle emission by a body.  For example, it is easy to show that Einstein's analysis adapted to the decay $\pi_0 \rightarrow e^+ + e^-$ leads us again to mass-energy equivalence.   However, this example, involving only material objects, makes explicit  the role of momentum conservation in obtaining $m_0 = L/c^2$ with now $L$ being the total energy of the emitted electron/positron pair in the rest frame of the pion of mass $m_0$.  Let $E_+$ and $E_-$ be the energies of the positron and electron in the rest frame of the pion, and let $p_+$ and $p_-$ be the magnitudes of their respective momenta in the pion's rest frame. Then, from momentum conservation, we obtain $p_+ = p_-$ which implies 
$\frac{m_e u_+}{\sqrt{1-\frac{u_+^2}{c^2}}} = \frac{m_e u_-}{\sqrt{1-\frac{u_-^2}{c^2}}}$ where $m_e$ is mass of the electron or positron and $u_+$ ($u_-$) is the speed of the positron (electron) in the rest frame of the pion.  From this equation we obtain $u_+ = u_- $ and hence $E_+ = \frac{m_e c^2}{\sqrt{1-\frac{u_+^2}{c^2}}}  = \frac{m_e c^2}{\sqrt{1-\frac{u_-^2}{c^2}}}  =E_-$.

Nothing in the analysis given by Einstein forbids us to apply his arguments to the emission of sound waves instead of radiation.  However, such an analysis might lead one n\"aively to predict massive amounts of mass being lost in the conversion into sound energy, due to the velocity of sound being so much smaller than $c$.  To avoid such nonsensical results, we  need to take into account momentum considerations.  Specifically, we use the fact that only zero energy phonons can transport momentum \cite{Kittel}. Thus it is impossible to assert anything about mass-energy equivalence using a process as the one considered by Einstein with sound emission instead of light emission, in spite of the fact that there is nothing in Einstein's paper which would prohibit one from doing such. 

In fact, neglecting possible general relativistic modifications as described in  \cite{Esposito}, the interaction of matter with sound waves should be considered, at least to first order in $v/c$, as a Galilean invariant theory,  for which conservation of momentum as a law involving  \underline{ponderable matter only} holds true, and which also respects the action/reaction principle.  Poincar\'e has made it quite clear in his Festschrift article \cite{Poincare} and elsewhere \cite{Poincare3} that only Galilean invariant theories with their presupposition of absolute simultaneity ($t'=t$) can be compatible with Newton's third law \cite{Zahar}.  There is no  analog of Eq. (1) for sound waves and sound does not transport momentum, nor does it have inertia.  

The sound analogy is very interesting from our standpoint.  For sound emission it is especially clear that we really have no right to single out the rest frame of the particle as the inertial frame where we declare the energies of the two sound waves moving in opposite directions to be the same. We see this as follows.  Sound waves do not carry momentum, so, to an observer in any inertial frame, the body loses energy, but never momentum nor mass. (Since momentum is proportional to mass, the body's mass cannot change, provided its momentum and velocity do not change.)   How much sound energy travels to the left and how much to the right depends upon the detailed dynamical make-up of the body emitting the sound waves, and it is easy to see from the Doppler formula for sound that we can always find an inertial frame where the energies of the two sound waves, the one to the left and the one to the right, are equal.  Furthermore, conservation of energy implies that the sum of the energies of the two sound waves is equal to the energy lost by the body.  This is exactly the situation described in Section 5 except it's for sound.  

Repeating Einsten's analysis for this case leads to $K_0 -K_1= (E'_+ - E_+)+
(E'_- - E_-)$.  Now, however, we must have $K_0 =K_1$, since the momentum and mass of the body do not change.  Hence $E'_+ + E'_- =  E_+  + E_-$. In other words, the sum of the energies of the two sound waves, and, hence, the energy lost by the body, are the same in all inertial frames, making it clear that nothing useful about mass-energy equivalence can be gotten out of arguments such as Einstein's in the case of sound. 
It also makes clear that Einstein's treatment cannot lead to $E=mc^2$ without attributing momentum and mass to radiation.   

Now to our simplified description of the Ives criticism adapted to pion decay. The decay  $\pi_0 \to \gamma + \gamma$ is depicted from different perspectives in Figs. 8, 9 and 10.  Fig. 8 shows the situation in the rest frame of the pion with the photon $\gamma_+$ having energy $E_{\g_+}$ emitted to the right and the photon $\gamma_-$ having energy $E_{\g_-}$ emitted to the left.  

Fig. 9 describes how the decay is seen by an observer moving to the left with speed $v$ relative to the pion.   In the rest frame $S'$ of the moving observer  the pion is seen to be moving to right with speed $v$.  The moving observer designates by $E'_{\g_+}$  and $E'_{\g_-}$ the energies of the two photons where the photon with energy $E'_{\g_+}$ moves to the right and the other one with energy $E'_{\g_-}$ moves to the left.

\begin{figure}[htb]
\begin{center}
\begin{tikzpicture}[
    media/.style={font={\footnotesize\sffamily}},
    wave/.style={
        decorate,decoration={snake,post length=1.4mm,amplitude=.6mm,
        segment length=2mm},thick},
    interface/.style={ postaction={draw,decorate,decoration={border,angle=-45,
                    amplitude=0.3cm,segment length=2mm}}},
    scale=.9] ]
\draw[->,wave]  
(-180:.4cm)--(-180:3.2cm)node[left]{$~$};
\node at (-3, 2) {$S$};
    \draw[->,wave]
       (0:.4cm)--(0:3.2cm)node[right]{};
\node at (-2,+.8) {$\gamma_-$};
\node at (0,+1) {$E_0$};

\node at (2.5,-.8) {$E_{\gamma_+}$};
\node at (-2.5,-.8) {$E_{\gamma_-}$};
\node at (2,+.8) {$\gamma_+$};
\node at (0, 0) {$\pi_0$};
\draw[line width=1pt](0,0)circle(.4cm);. 
\end{tikzpicture}  

\caption{Decay of pion  into two photons as viewed in the rest frame $S$ of the pion.  The photon with energy $E_{\g_{+}}$ moves to the right and the other one with energy $E_{\g_{-}}$ moves to the left. The rest frame energy of the pion is $E_{0}$.}
\end{center}
\end{figure}
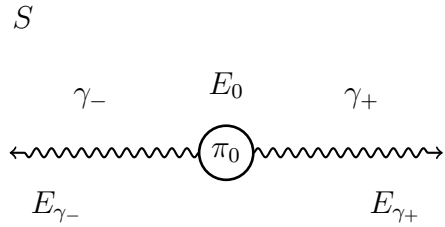

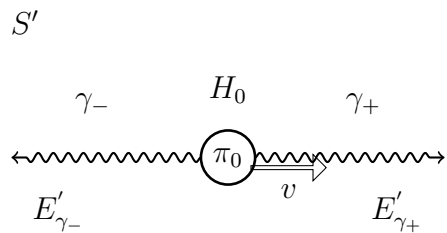
\begin{figure}[htb]
\begin{center}
\begin{tikzpicture}[
    media/.style={font={\footnotesize\sffamily}},
    wave/.style={
        decorate,decoration={snake,post length=1.4mm,amplitude=.6mm,
        segment length=2mm},thick},
    interface/.style={postaction={draw,decorate,decoration={border,angle=-45,
                    amplitude=0.3cm,segment length=2mm}}}, scale=.9]
 \draw[->,wave]  
(-180:.4cm)--(-180:3.2cm)node[left]{$~$};
\node at (.88,-.15) [single arrow,draw,inner sep=.3,minimum width=.35cm,minimum height=1cm,single arrow head extend=0.1cm] {};
\node at (-3, 2) {$S'$};
\node at (.9, -.5) {$v$};
\draw[->,wave]
       (0:.4cm)--(0:3.2cm)node[right]{$~~$};
\node at (-2,+.8) {$\gamma_-$};
\node at (0,+1) {$H_0$};
\node at (2.5,-.8) {$E_{\gamma_+}^{'}$};
\node at (-2.5,-.8) {$E_{\gamma_-}^{'}$};
\node at (2,+.8) {$\gamma_+$};
\node at (0, 0) {$\pi_0$};
\draw[line width=1pt](0,0)circle(.4cm);
 
\end{tikzpicture}  

\caption{Decay of pion  into two photons as viewed in the reference frame $S'$ where the pion moves to the right with speed $v$.  The photon with energy $E'_{\g_{+}}$ moves to the right and the other one with energy $E'_{\g_{-}}$ moves to the left. The energy of the pion in $S'$ is $H_0$.}

\end{center}
\end{figure}

Finally we consider in Fig. 10 the situation in an arbitrary frame $S''$ of an observer moving with a velocity $-\d \vec v$ relative to the pion, with $\d$ being as in Section 5.  In this frame the pion is seen to be moving to the right with speed $\d v$.   

\begin{figure}
\begin{center}
\begin{tikzpicture}[
    media/.style={font={\footnotesize\sffamily}},
    wave/.style={
        decorate,decoration={snake,post length=1.4mm,amplitude=.6mm,
        segment length=2mm},thick},
    interface/.style={postaction={draw,decorate,decoration={border,angle=-45,
                    amplitude=0.3cm,segment length=2mm}}}, scale=.9]
\draw[->,wave]  
(-180:.4cm)--(-180:3.2cm)node[left]{$~$};

\node at (.88,-.15) [single arrow,draw,inner sep=.3,minimum width=.35cm,minimum height=1cm,single arrow head extend=0.1cm] {};
\node at (-3, 2) {$S''$};
\node at (.9, -.5) {$\d v$};
    \draw[->,wave]
       (0:.4cm)--(0:3.2cm)node[right]{};
\node at (-2,+.8) {$\gamma_+$};
\node at (0,+1) {$E''_0$};
\node at (2.5,-.8) {$E_{\gamma_-}^{''}$};
\node at (-2.5,-.8) {$E_{\gamma_+}^{''}$};
\node at (2,+.8) {$\gamma_-$};
    \node at (0, 0) {$\pi_0$};
\draw[line width=1pt](0,0)circle(.4cm);
 \end{tikzpicture}

\caption{Decay of pion  into two photons as viewed in the arbitrary frame $S''$, where pion moves to the right with speed $\d v$ and has energy $E''_0$.  The photon with energy $E''_{\g_+}$ moves to the right and the other one with energy $E''_{\g_-}$ moves to the left.}
\end{center}
\end{figure}
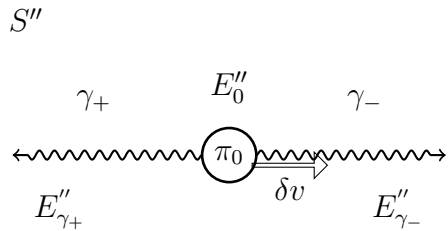

It should be clear from the previous section, how we shall proceed to explain the logic of the Ives criticism.  Without the justification provided by momentum conservation for equating the energies of the two photons in the rest frame of the pion, we should take the arbitrary $S''$ frame of Fig. 10 as the frame where we  declare the two energies $E''_{\g_+}$ and $E''_{\g_-}$ equal, since all inertial frames must be treated equally by the relativity principle.  So we equate the  energies of the two photons in the $S''$ frame.  With $L=  E_{\g_+}+E_{\g_-}$, conservation of energy in $S$ and $S'$ give
$$ E_0 = E_{\g_+} + E_{\g_-} = L \eqno(12^\textrm{quater})$$
and 
$$ H_0 =  E'_{\g_+} + E'_{\g_-}  = \frac{L\left(1- \frac{\d v^2}{c^2} \right)}{\sqrt{1-\frac{v^2}{c^2}}}, \eqno(13^\textrm{quater})$$
respectively. (Compare Eqns. (12$^\textrm{ter}$) and (12$^\textrm{ter}$).)
From Eqns. (12$^\textrm{quater}$)  and (13$^\textrm{quater}$), we obtain
$$ H_0 -E_0 = L\left\{\frac{\left(1- \frac{\d v^2}{c^2} \right)}{\sqrt{1-\frac{v^2}{c^2}}}-1 \right\}  \eqno(45)$$
which is relativistically exact.

Now, we know from special relativity, that the relativistically exact expression for the kinetic energy of a pion of mass $m_0$ in the moving frame $S'$ is
$$ K_0^\textrm{pion} = m_0 c^2 \left\{\frac{1}{\sqrt{1-\frac{v^2}{c^2}}} -1 \right\} . \eqno(46)$$
Comparing Eqns. (45) and (46) leads us to  the equation
$$ H_0 - E_0 = \frac{L\left\{\frac{\left(1- \frac{\d v^2}{c^2} \right)}{\sqrt{1-\frac{v^2}{c^2}}}-1 \right\}}{ m_0c^2 \left\{\frac{1}{\sqrt{1-\frac{v^2}{c^2}}} -1 \right\}}   K_0^\textrm{pion} .\eqno(47)$$   

Now we come to Einstein's \textit{assumptions}, which were criticized by Planck and Ives, i.e. our Eqns. (4) and (5).  For our special case of pion decay they reduce to
$$H_0 - E_0 = K_0 + C \eqno(4')$$
$$0 - 0 = 0 + C . \eqno(5')$$
Hence $C = 0$, either by Eq. (5$'$) or by the Proposition, and thus $$H_0 -E_0 = K_0 \eqno(48)$$
 Combining Eqns. (47) and (48)  gives 
$$ K_0 = \frac{L\left\{\frac{\left(1- \frac{\d v^2}{c^2} \right)}{\sqrt{1-\frac{v^2}{c^2}}}-1 \right\}}{ m_0c^2 \left\{\frac{1}{\sqrt{1-\frac{v^2}{c^2}}} -1 \right\}}  K_0^\textrm{pion} .\eqno(49)$$
The nonrelativistic limit of Eq. (46) gives $$K_0^\textrm{pion} = \halb m_0 v^2 \eqno(50) $$ to order $v^2/c^2$.  

For $K_0$ the situation is more  complicated.  Eq. (49) makes clear that Einstein is not right in asserting, solely on general grounds, that the difference between the energies $H_0$ and $E_0$  of a particle moving with velocity $\vec v$ and at rest, respectively, is the kinetic energy of the particle, which, in our case, is the pion.  Without momentum considerations, we can only assert that $K_0= H_0 -E_0$ is proportional to the kinetic energy of the pion.  Something else is needed in order to set the scale for uniquely fixing the particle's inertia, and that, of course, is conservation of momentum along with momentum of radiation.  Without such arguments, we can only assert that $$K_0 = \halb m_\d \, v^2 \eqno(51)$$  to order $v^2/c^2$ where $m_\d$ is a number, which is proportional to $$ m_0 = \frac{E_0}{c^2}  = \frac{L}{c^2}.  $$ 
In fact, performing a power series expansion of the right hand side of Eq. (49) in powers of $v^2/c^2$ and making use of the nonrelativistic approximations for $K_0^\textrm{pion}$ and $K_0$ given by Eqns. (50) and (51) leads us to
$$ \halb m_\d v^2 = \frac{L}{c^2} (1-2\d) v^2 $$
and thus
$$ m_\d = \frac{L}{c^2} (1-2\d) . \eqno(52)$$
Applying the same nonrelativistic analysis to Eq. (46) leads us to  
$$ m_0 = \frac{L}{c^2}. \eqno(53)$$
Thus from Eqns. (52) and (53) we get
$$m_\d= (1-2\d) m_0 \eqno(54).$$

Without further considerations, such as momentum conservation, $\d$ must be taken to be arbitrary, and, if this is the case, then $K_0 =H_0 - E_0$, i.e. our Eq. (4$'$), with $K_0=K_0^\textrm{pion}$ can only be true \textit{if and only if} that which he derived,  i.e. $E_0 = m_0 c^2$,  is true.  We can summarize our arguments regarding the logical fallacy of Einstein's analysis as follows:

\noindent
\textit{Theorem 1:}  \textit{Assuming no distinguished priority is given to any one particular inertial frame, and \underline{using only energy considerations} such as conservation of energy for the pion decay and the relativistic Doppler formula for the light pulses, then, given that relativistic quantities such as  kinetic energy have well-defined nonrelativistic limits, we have with $$K_0 = H_0 -E_0$$ 
that
$$  K_0 =  K_0^{\textrm{pion}}  \quad \iff \quad  E_0 = m_\d c^2   \quad \iff \quad  \d = 0  $$
where $K^\textrm{pion}_0$ is the relativistic kinetic energy of the pion and where $$m_\d := \lim\limits_{c^2 \rightarrow \infty}\frac{K_0}{v^2/2} $$}
with $\frac{E_0}{c^2}$ held fixed (cf. footnote 3). 

\noindent
Proof:  $K_0 =  K_0^{\textrm{pion}}$  $\Rightarrow$ $m_\d = \lim\limits_{c^2 \rightarrow 0} \frac{K_0}{v^2/2} = \lim\limits_{c^2 \rightarrow 0} \frac{K_0^{\textrm{pion}}}{v^2/2} = m_0$, by Eq. (46).  Thus $m_0=m_\d$ and  hence by Eq. (53) and $E_0 =L$ we have $$E_0= L=m_0 c^2 = m_\d c^2.$$  

Now to show $E_0 = m_\d c^2 \Rightarrow \d =0$:  from $L= E_0 = m_\d c^2$ together with Eq. (53) we obtain
$$m_0= \frac{E_0}{c^2} =  \frac{m_\d c^2}{c^2} = m_\d .$$
But we also have $m_\d = (1-2\d) m_0$ by Eq. (54).  Thus $\d = 0$.

Finally, conservation of energy in $S$ and $S'$ i.e. $E_0 = L$ and $H_0 = L'$, and the relativistic Doppler formula, led us to Eqns. (12$^\textrm{quater}$) and (13$^\textrm{quater}$) for $E_0$ and $H_0$, respectively.  Using these equations we get
$$K_0= H_0-E_0= E_0 \left\{\frac{\left(1- \frac{\d v^2}{c^2} \right)}{\sqrt{1-\frac{v^2}{c^2}}}-1 \right\}  \eqno(45'). $$ 
Using $\d=0$ this becomes
$$  K_0= H_0-E_0 =E_0  \left\{\frac{1}{\sqrt{1-\frac{v^2}{c^2}}}-1 \right\} =  K_0^{\textrm{pion}},$$ since $E_0 = m_0c^2$ by Eq. (53) and $E_0=L$.   

We also have for the more general situation considered in sections 3 and 5 (of which pion decay is a special case) the following result. 

\noindent
\textit{Theorem 2:} \textit{With the same general assumptions as in Theorem 1, i.e.  assuming no distinguished priority is given to any one particular inertial frame, and using only energy considerations such as conservation of energy and Einstein's assumptions, then   $\d =0$ $\iff$  $E_{\g_+}= E_{\g_-}$. }

\vskip .2cm

\noindent
Proof: ($\Rightarrow$) $\d=0$ implies by Eq. (37) that  $E_{\g_+}= E_{\g_-}$, since for $\d=0$, $E_{\g_\pm}= E''_{\g_\pm}$ and $E''_{\g_+}= E''_{\g_-}$ by assumption.

\vskip .2cm

\noindent
($\Leftarrow$)    
Using  $E_{\g_+}= E_{\g_-}$ in Eq. (37) gives $$E''_{\g_+}\left(\frac{2 \d v}{c} \right) = E''_{\g_-}\left(\frac{2 \d v}{c} \right)  =0,$$ 
since $E''_{\g_+}= E''_{\g_-}$.  This implies $\d=0$.   (Note that $\d=0$ in Eq. (17$^\textrm{ter}$) gives Einstein's result $m_0 - m_1 = \frac{L}{c^2} $.)

Einstein has no right to assume that $H_0 -E_0 = K_0 +C$,  with $K_0$ being the \textit{relativistic kinetic energy of a body of mass} $m_0$,  any more than he any right to declare  $E_{\g_+} = E_{\g_-} = \halb L$, without providing some additional justification than what is in his paper.   These are the two oversights in Einstein's paper. 

Einstein does have the right, based on his general argumentation relating energies in different reference frames to kinetic energy, to declare $H_0 -E_0$ equal to the relativistic kinetic energy of a particle of \textit{some mass}, but not of mass $m_0$ i.e. the mass of the pion, without additional justification. It is clear from any serious study of his paper, that this is the point which Ives tried to make, and it is this point that Field and Stachel and Torretti fail to see or choose not to see.  

Stachel and Torretti  concede  that ``Had he used"  Eq. (46) for $K_0$ in $H_0 -E_0 = K_0 +C$, then Einstein ``could indeed have been justly accused of question begging" \cite{Stachel}.  Stachel and Torretti  view $H_0-E_0$ as an arbitrary function of $m_0$ and $v^2$ which  has limit $\halb m_0 v^2$ for small $v$,  assuming the limit exists.  But this is tantamount to assuming  $\d=0$ and hence $m_\d={L}/{c^2} = m_0$.  We see this as follows. The correct equation for $K_0$ is Eq. (45) since $K_0=H_0-E_0$. It has as its limit, as $c\rightarrow \infty$ $$\lim\limits_{c \rightarrow \infty} K_0 =\lim\limits_{c \rightarrow \infty} (H_0 - E_0)= \halb \frac{L{(1-2 \d)}}{c^2} v^2 = \halb{m_0}{(1-2 \d)} v^2 , $$ 
since $\frac{L}{c^2} = m_0$ by Eq. (53).  This reduces to  $$\halb m_0 v^2$$ if and only iff $$ \d=0 \qquad \iff \qquad H_0 - E_0  = K_0 $$
with $K_0$ given by Eq. (46) i.e. with $K_0 = K_0^\textrm{pion}$.  So,  from our viewpoint, \textit{Einstein has used Eq. (46) for} $K_0$ \textit{ in} $H_0 -E_0 = K_0 + C $, and, hence, by Stachel and Torretti's own admission, he is guilty of question begging.

   We conclude with a brief description of Ives' paper.  It should make it clear how that which we have just described is the Ives' criticism of Einstein's paper specialized to the simple situation considered by us, i.e. of pion decay.  After a short  introduction where Ives distinguishes between two versions of $E=mc^2$, one of which he attributes to Poincar\'e, which are $E =m_Rc^2$ and $E= m_Mc^2$, with $m_R$ and $m_M$ being the ``mass of radiation" and the ``mass of matter," respectively, Ives starts out with explaining in relativistically exact terms Poincar\'e's derivation of mass-energy equivalence based on conservation of energy and \textit{conservation of momentum}, together with Poincar\'e's momentum of radiation.  It is clear from what he writes that Ives is fully aware of the essential role  momentum conservation plays in any legitimate derivation of $E=mc^2$ and in his summary,  he emphasizes this point.  

Finally, Ives  presents his analysis of the Einstein paper:  Ives makes use of Eq. (46) twice to obtain the following relativistically exact expression for the difference in the kinetic energies of the particle of Fig. 3 before and after the emission of radiation
$$ K_0 - K_1 = (m_0 -m_1) c^2 \left\{ \frac{1}{\sqrt{1-\frac{v^2}{c^2}}} -1 \right\} . \eqno(55)$$ 
On the other hand, from Eqns. (16) and (14) and (15) we obtain
$$ (H_0 -H_1) - (E_0 - E_1) =  L \left\{  \frac{1}{\sqrt{1-\frac{v^2}{c^2}}} -1 \right\} . \eqno(56)$$
Now using Einstein's equations, i.e. our Eqns. (14) and (15), we get 
$$(H_0 - E_0) -(H_1-E_1) = K_0 -K_1 . \eqno(57)$$
By comparing Eqns. (56) and (57)  and making use of Eq. (55) Ives is led to
$$  (H_0 - E_0) -(H_1-E_1)  = \left\{ \frac{L}{(m_0-m_1)c^2} \right\} (K_0-K_1) .$$
It is clear based on Einstein's argumentation about kinetic energy in terms of rest energy and energy while in motion that this can be split up into the two equations
$$ (H_0 - E_0) = \left\{ \frac{L}{(m_0-m_1)c^2} \right\} (K_0 +C) $$
and 
$$ (H_1 - E_1) = \left\{ \frac{L}{(m_0-m_1)c^2} \right\} (K_1 +C) . $$
(In fact, according to the Proposition, we can even take $C=0$.) 
Ives then emphasises that these equations are not Eqns. (14) and (15), rather, instead, as we have explained in much more detail, they differ from them by the proportionality factor 
$$\left\{ \frac{L}{(m_0-m_1)c^2} \right\} .$$ 
They agree with Eqns. (14) and (15) if and only if this proportionality factor is set equal to one, which means 
$$ L= {(m_0-m_1)c^2} .$$
But this is exactly what Einstein set out to prove by using Eqns (14) and (15).\footnote{Incidentally, Field is wrong in attributing to Ives the statement that Einstein was guilty of the logical error of \textit{petitio principii} (cf. Section 5 of Ref. \cite{Field}).   It came from Max Jammer, not from Herbert Ives.  Jammer, having studied the Ives' criticism and agreeing with it, wrote \cite{Jammer}:   ``\textit{It is a curious incident in the history of scientific thought that Einstein’s own derivation of the formula $E = mc^2$, as published in his article in the {\textsl{Annalen der Physik}}, was basically fallacious. In fact, what for the layman is known as the most famous mathematical formula ever projected  in science was but the result of petitio principii, the conclusion of begging the question.}"   Ives, on the other hand,  only wrote \cite{Ives}: ``\textit{What Einstein did by setting down these equations [our Eqns. (14) and (15)] (as ``clear") was to introduce $$L/(m-m')c^2 =1.$$ 
Now this is the very relation the derivation was supposed to yield.}" }

More specifically, following the analysis of sections 3 and 5 that led us (and Einstein) to Eq. (16) but without assuming  $E_{\g_{+}}= E_{\g_-}$,  we get, by using Eqns. (38) together with $K_0 = H_0 - E_0$ and $K_1= H_1 - E_1$  that
$$ K_0 - K_1 = L \left\{  \frac{1}{\sqrt{1-\frac{v^2}{c^2}}} -1 \right\} + \frac{1}{\sqrt{1-\frac{v^2}{c^2}}} \left( \frac{E_{\g_{+}}}c -  \frac{E_{\g_{-}}}c \right) v  \eqno(58)$$
where $L = E_{\g_{+}} + E_{\g_-}$.     Now, according to Eq. (37) 

$$  E_{\g_\pm} = \frac{E''_{\g_\pm}\left(1 \mp\frac{\d v}{c}\right)}{\sqrt{1- \frac{(\d v)^2}{c^2}}} , \eqno(59)$$
so that
$$\frac{E_{\g_{+}}}c -  \frac{E_{\g_{-}}}c = -\frac{\d v}{c^2} \frac{( E''_{\g_+} + E''_{\g_-} )} {\sqrt{1- \frac{(\d v)^2}{c^2}}} = -\frac{\d v}{c^2} L , \eqno(60)$$
since $ E''_{\g_+} = E''_{\g_-}$.  Substitution of this result into Eq. (58) gives
$$ K_0 - K_1 = L \left\{  \frac{1}{\sqrt{1-\frac{v^2}{c^2}}} - \frac{\d v^2}{c^2}-1 \right\} . \eqno(61)$$  
Now Einstein's nonrelativistic approximation argument using a Maclaurin expansion of the first term of the right hand side of this equation leads to 
$$ \halb (m_0 - m_1)v^2 = \halb \frac{L}{c^2} (1-2 \delta) ,$$ 
which gives Einstein's desired conclusion, i.e. $L = (m_0 -m_1) c^2$,  if and only if $\d=0$, confirming part of Theorem 1,  namely $\d=0$ if and only if $L = (m_0 - m_1)c^2$.  Furthermore, from Theorem 2, we get $ E_{\g_+} = E_{\g_-}$ if and only if $\d=0$.   In summary, \textit{if we don't want to make use of momentum conservation and momentum of radiation, then we have to make use of $L = (m_0 -m_1) c^2$ in order to legitimately obtain $L = (m_0 -m_1) c^2$ out of Einstein's argumentation.}

At first glance, it might seem that in order for Eq. (58) to make sense in the nonrelativistic limit, we must set $E_{\g_{+}} =  E_{\g_-}$, since the  left hand side of Eq. (58) in the nonrelativistic limit consist of kinetic energy terms which are proportional to $v^2$, nonrelativistically, and, therefore, there should not be any linear terms in $v$ on the right hand side of Eq. (58) as is the last term. This would mean that the last term in Eq. (58) must vanish for arbitrary $v$, implying that  we must have $E_{\g_{+}} =  E_{\g_-}$.  However, the analysis leading us to Eq. (61) shows that this is not quite correct, since $E_{\g_{\pm}}$ according to Eq. (59) actually involves contribution from a term linear in $v$, i.e. the term $\pm \d v/c$ in the numerator, so that the last term of the right hand side of Eq. (58) does contribute a correction term of order $v^2$ to the right hand side in the nonrelativistic limit.  

On the other hand, to lowest order in $v/c$ we do have $E_{\g_{\pm}} = E''_{\g_{\pm}}$, so that if were to replace $E_{\g_{+}}$ and $E_{\g_-}$ in Eq. (58) by   $E''_{\g_{+}}$ and $E''_{\g_-}$, we can possibily interpret what Planck might have meant by insisting that Einstein's argument was ``permissible only as a first approximation" \cite{Planck}.  In this case, the last term vanishes and Eq. (61) becomes 
$$ K_0 - K_1 =  \widetilde  L \left\{  \frac{1}{\sqrt{1-\frac{v^2}{c^2}}} -1 \right\}  \eqno(61')$$ 
where $\widetilde{L} = E''_{\g_+} + E''_{\g_-}$.  Taking the nonrelativistic limit of this equation we get
$$\halb (m_0-m_1) v^2 = \frac{ \widetilde{L} } {c^2} $$
which is Einstein's desired conclusion from which his result on mass-energy equivalence follows. However, $\widetilde{L}$ is not the total energy $L$ of radiation, rather we have by Eq. (59) 
$$ L = \frac{{\widetilde{L}}}{\sqrt{1-\frac{(\d v)^2}{c^2}}} .$$
In this sense, Einstein's analysis, as Planck asserts, could be considered approximately valid to some crude lowest order of approximation.  We leave it to the reader to convince himself that what we have just explained is essentially the Ives' interpretation of the Planck criticism (cf. \cite{Ives}, p. 543).

\section{Conclusions}
In in his recent E.J.P. article \cite{Field}, J. H. Field, makes a bold  attempt to vindicate Einstein's 1905 article from  all wrong by disposing of all serious criticisms of the paper, including those of authors Ives, Ohanian and Hecht.  Specifically, he writes: ``Claims in the literature that Einstein's analysis was flawed, lacked generality, or was not rigorous, are rebutted."  Needless to say, our results contradict his findings with regards to the Ives criticism.

Einstein's argument to get at $E=mc^2$, based strictly on what is written there, i.e. using only the relativity principle, energy considerations and his relativistic Doppler formula, is lacking and cannot be considered as sufficient.  It suffers from the error of circular reasoning in the sense that $K_0=H_0-E_0$ together with energy conservation implies $E_0=m_0c^2$ only for the special case $\d=0$.  But $\d=0$ if and only if $E_0=m_0c^2$.   The route Einstein followed is clearly not the way to proceed.  Rather, one should, in addition to conservation of energy, \textcolor{black}{make} use of the other conservation law used in collision and emission processes, which is that of momentum conservation. 

In defense of Einstein's paper we could assume that, by declaring $E_{\g_{+}} = E_{\g_-}$ at the onset of his considerations, he had implicitly made use of momentum conservation and momentum of radiation, but this would imply that his derivation should be viewed, at best, only as a streamline proof of $E=mc^2$ based on the more fundamental works of Poincar\'e \cite{Poincare} and Hasen\"ohrl \cite{Hasenoehrl}.\footnote{In fact, it may well be that this is how Field actually interprets the situation!  For he writes: ``Whether a philosopher would consider this derivation ‘rigorous’ is
perhaps an open question but I submit, in agreement with Stachel and Torretti [11] and Fadner
[9], and contrary to Planck [3], Ives [5] and recent assertions of Hecht [7] and Ohanian [6],
that most physicists would."}  After all, it is hard to imagine that Einstein, at the time he wrote his paper, was not aware of Poincar\'e's Festchrift article,  which was one of the most important and widely read physics papers of that time \cite{Martinez}, and it seems almost certain he would have been aware of Hasen\"ohrl's papers published some months before in the same journal to which he submitted his first two relativity papers \cite{Rothman}.  

On the other hand, as I explained at the end of Section 5, such a view seems to conflict with what Einstein wrote in his paper, suggesting that he  might have thought it unecessary to atrribute momentum to radiation, a very serious error!   For this to be true presupposes that he would have been unaware of Poincar\'e's and Hasen\"ohrl's papers.   In any case, it seems worthwhile to see how Einstein's treatment of his emission process would have gone had Einstein also made use of momentum conservation in addition to conservation of energy, something to which I now turn.

If we follow Einstein's analysis, but without assuming at the onset $E_{\g_{+}} = E_{\g_-}$, we get from Eqns. (38) together with $K_0 = H_0 - E_0$ and $K_1= H_1 - E_1$  that 
$$ K_0 - K_1 = L \left\{  \frac{1}{\sqrt{1-\frac{v^2}{c^2}}} -1 \right\} + \frac{1}{\sqrt{1-\frac{v^2}{c^2}}} \left( \frac{E_{\g_{+}}}c -  \frac{E_{\g_{-}}}c \right) v ,  $$
since, from conservation of energy, $E_0 -E_1= L = E_{\g_{+}} + E_{\g_-}$ and $H_0 -H_1= E'_{\g_{+}} + E'_{\g_-}$.
 
Conservation of momentum in the $S$ frame together with Poincar\'e's momentum of radiation gives 
$$ E_{\g_{+}} = E_{\g_-}  .$$
From momentum conservation in the $S'$ frame and Eq. (38)  we get  
$$p_0 - p_1 = \frac{E'_{\g_{+}}}c -  \frac{E'_{\g_{-}}}c = \frac{L}{c^2} v $$
where $p_0$ and $p_1$ are the relativistic momenta of the particle before and after emission of radiation, respectively. Now, following Einstein, use a nonrelativistic approximation of this equation to obtain to lowest order in $v/c$
$$m_0 v - m_1 v =  \frac{L}{c^2} v $$
from which we we obtain Einstein's desired result i.e. $m_0 -m_1 =  \frac{L}{c^2}$.  Furthermore, if we multiply both sides of this equation by $\halb v$ we get
$$ K_0 - K_1 =\halb m_0 v^2 - \halb m_1 v^2 = \halb \frac{L}{c^2} v^2 $$
which, if we neglect the middle term, is the final equation of Einstein's paper.

Finally, I would like to briefly comment on Fadner's peculiar objection to Ives' criticism based on what Fadner calls ``the logical deductive development of a scientific theory" \cite{Fadner}.   According to our viewpoint, this seems entirely irrelevant, since the matter  is not one involving the interrelationship between $n$ postulates whose truth and meaning are unambiguous and without dispute.  Rather, it is a matter of confusion in interpretation of Einstein's assumptions.   As I have tried to make abundantly clear in this paper, Einstein's has no right to interpret $K_0$ as the relativistic kinetic energy of a particle of inertial mass $m_0$, but rather only the relativistic kinetic energy associated with some mass, unless, of course, he assumes that which he wants to prove, or invokes additional considerations involving momentum. The correct interpretation of $K_0$ in Einstein's treatment should be that of a kinetic energy associated with a mass which depends upon a parameter $\d$, and, according to Theorem 1, $K_0$  becomes the kinetic energy of a particle of mass $m_0$ if and only if $\d =0$ which is equivalent to $E_0 =m_0 c^2$.  

In conclusion, my elaboration of the logic of Ives' critisicm of Einstein's paper presented here has as its aim to make very precise how Einstein's oversight in neglecting momentum considerations, specifically momentum of radiation, leads to ambiguity in his paper regarding the interpretation of kinetic energy.  It is this ambiguity in interpretation of the kinetic energy of the body on which both Planck and Ives base their criticisms.  Those who fail to see or choose not to see that there is a problem with Einstein's neglect of conservation of momentum and momentum of radiation are free to do so, but failing to see it, or choosing not to see it, does not invalidate the problem. Both Poincar\'e \cite{Poincare} and Hasen\"ohrl \cite{Hasenoehrl}, \cite{Hasenoehrl1} made momentum considerations central in their treatements of mass-energy equivalence.  

Stachel and Torretti in their paper apparently take it for granted that Einstein implicitly used momentum conservation, for they write that the body in its rest frame ``loses energy but not momentum" \cite{Stachel}. Using momentum conservation one is led to a unique interpretation of $K_0$ as the kinetic energy of a particle of mass $m$, in which case the basis on which the Ives objection rests is no longer valid.   This is the reason why they can contend that ``the paper [Einstein's] was basically sound" \cite{Stachel} and that ``we have to declare that Ives, Jammer, and Arzeli\`es--not Einstein--are guilty of a logical error" \cite{Stachel}.    In other words, they corrected Einstein's oversite by sneaking the premise of momentum conservation into Einstein's argument, so that they could get around the Ives criticism.

However, they are silent as to what might have led Einstein to implicitly assume momentum conservation, nor do they explain how he could have made implicit use of momentum conservation, even though Einstein never made any mention whatsoever of the word momentum or of momentum conservation in his paper.  Such would mean that he would have had to attribute momentum to radiation, something which was far from notrivial at beginning of the twentieth century.  Nor do they provide any information on important questions, such as how much momentum is carried off by the radiation and the relation between the energy and the momentum of radiation.  Such questions are key concerns in the paper of Poincar\'e, and they also should have been important questions for Einstein.  Thus, even if one chooses to believe that Einstein had implicitly made use of momentum conservation and momentum of radiation in his paper, without making them explicit, Einstein's derivation of $E=mc^2$ is not completely convincing.  Furthermore, in such a case, the Ives's criticism cannot be viewed as wrong, rather only as irrelevant or not valid, since in this case the basis on which the Ives objection rests is no longer true. 

Ohanian apparently fell victim to Stachel and Torretti's ploy in taking it for granted that Einstein made implicit use of momentum conservation.  He offers no objection to Einstein assuming  that the ``body emits two pulses of light of energy $E/2$ in opposite directions in its rest frame"  \cite{Ohanian}.  This explains why he can agree with Stachel and Torretti's findings.  As mentioned in the introduction, a proper understanding of the logic of the Ives criticism can only be understood in terms of Einstein's failure to take into account momentum conservation.  

Field, on the other hand, seems to be fully aware of the importance of momentum considerations and momentum of radiation in any correct analysis of Einstein's emissison process, but he seems to think that such an approach would have been ``unknown to Einstein at the time of writing" \cite{Field}.  Apparently, he thinks that Einstein would have had to have used the exact relativistic formula for momentum, which was not yet known, in order to obtain $E=mc^2$.  But this is not the case.   Just as Einstein used the nonrelativistic approximation for kinetic energy, so also there is no reason why he could not have used the nonrelativistic formula for momentum, exactly as we did at the beginning of these conclusions, and as Poincar\'e did in his Festschrift paper.  

Finally, in spite of the fact that he seems to be fully aware of the importance of momentum considerations in getting at $E=mc^2$, Field apparently insists that it is possible to derive Einstein's result only by ``imposing energy conservation and the validity of Newtonian kinematics in the $\beta \; [=(v/c)] \rightarrow 0$ limit (both very weak postulates)" \cite{Field}.  Our analysis makes it clear that in this he is mistaken. 

In summary, it is necessary to give Einstein a lot of leeway in order to agree with Stachel and Torretti that his 1905 paper  on $E=mc^2$ is ``basically sound" \cite{Stachel}.   Specifically, we have to assume that he implicity borrowed from Poincar\'e his generalization of the law of momentum conservation to include radiation which means endowing radiation with momentum and inertia.  Such results were highly nontrivial and revolutionary at that time and without assuming that Einstein had made implicit use of them in his paper, it is seriously flawed.    This is the point which Ives tried to make and it is the reason for this exposition on the logic of the Ives paper.  To think otherwise is a misrepresentation of Ives' criticism.

\section*{Appendix: Proof of the Proposition}

Let us consider as in  Section 6 the situation of particle annihilation into radiation.  Such a decay can be viewed as the decay of a particle with nonzero energy into a particle with zero energy and zero momentum in all inertial frames, which means $E_1$ and $H_1$ are zero.  Now $\lim\limits_{c\rightarrow \infty}K_1= \halb \vec v \cdot \vec p_1$, so that with $\vec p_1 = 0$ we obtain $\lim\limits_{c\rightarrow \infty}K_1=0$.  Thus, by taking limits $c\rightarrow \infty$ of both sides of 
\[ H_1 -E_1 = K_1 + C ,\]
which is (Eq. 15), we get
\[ 0 - 0 = 0 + C.\]
Hence $C=0$.  Now, according to Einstein, for any body of rest energy $E$ and kinetic energy $K$ ``the difference
$H-E$ can differ from the kinetic energy $K$ of the body," with respect to the
system $S'$, ``only by an additive constant $C$, which depends on the choice of
the arbitrary additive constants of the energies $H$ and $E$" and ``\underline{does not change during the emission of light}" \cite{Einstein}.  This means that $C$ must be the same for the pion as for the particle (the vacuum) with zero energy and zero momentum into which it decays.  So we must have $C=0$ in Eq. (14) also.


\section*{References}
\begin{thebibliography}{50}
\bibitem{Poincare} Poincar\'e H  1900 La Th\'eorie de Lorentz et le Principe de R\'eaction \textit{Archives Néerland. Sci. exactes et naturalles} 2$^e$ \textit{s\'erie} {\bf 5} 252-278.
\bibitem{Einstein} Einstein A (1905) Ist die Tr\"agheit eines K\"orpers von seinem Energieinhalt 
abh\"angig?  \AP {\bf 18} 639- 641.

\bibitem{Planck} Planck M 1907 Zur Dynamik bewegter Systeme  \textit{Sitzungsberichte
der K\"oniglich-Preussischen Akad. der Wissenschaften} {\bf 29} (Erster Halbband)   542-570.
\bibitem{Field} Field J H 2014 \EJP {\bf 25} 123-126

\bibitem{Ives} Ives H E 1952 \JOSA 42 (8) 540-543.
\bibitem{Jammer} Jammer M (1961) \textit{Concepts of Mass in Classical and Modern Physics} (Mineola, NY: Dover).
\bibitem{Miller} Miller A I 1981 \textit{Albert Einstein's Special Theory of Relativity} (Reading, MA: Addison-Wesley)  87.
\bibitem{Arzieles} Arzeli\'es H 1966  {\it Rayonnement et Dynamique du Corpuscule Charg\'e Fortement Accel\'er\'e} (Paris: Gauthier-Villars) 74-79.


\bibitem{Stachel} Stachel J Torretti R 1982 \textit{Am. J. Phys.} {\bf 50} (8), 760-763.
\bibitem{Jammer2}  Jammer M 2009  \textit{Concepts of Mass in Contemporary Physics and Philosophy} (Princeton, NJ: Princeton University Press).

\bibitem{Ohanian}  Ohanian H C  2009 \textit{Studies in History and Philosophy of Modern Physics } {\bf 40} 167–173
\bibitem{Mermin} Mermin N D 2011  \textit{Studies in History and Philosophy of Modern Physics} {\bf 42} 1–2.
\bibitem{Ohanian2}  Ohanian H C 2012 \textit{Studies in History and Philosophy of Modern Physics} {\bf 43} 215–217 [In Ohanian's own words on the counterattacks: ``\dots as a warning to the unwary  (‘‘Here be dragons’’) \dots."]
\bibitem{Hecht} Hecht E 2011 \textit{Am. J. Phys.} {\bf 79} 591-600.

\bibitem{Abraham1} Abraham M 1902 Dynamik des Elektrons  \textit{K\"onigliche Gesellschaft der
Wissenschaften zu G\"ottingen: Mathematisch-physikalische Klasse. Nachrichten} 20–41.

\bibitem{Abraham2} Abraham M 1903 Prinzipien der Dynamik des Elektrons \AP {bf 10} 105–179.
\bibitem{Griffiths} Griffiths D J 2012 \textit{Am. J. Phys.} {\bf 80} 7-18.
\bibitem{Whittaker} Whittaker E T 1973 \textit{A History of the Theories of Aether and Electricity} (New York: Humanities Press)
\bibitem{Poynting} Poynting J 1884 On the Transfer of Energy in the Electromagnetic Field 
\PTRS \textit{London} {\bf 175} 343–361.
\bibitem{Thomson} Thomson J J 1881 {\textit{Phil. Magazine}} Series 5 {\bf{11}} 229-249.
\bibitem{Heaviside}  Heaviside O 1889 {\textit{Phil. Magazine}} Series 5 {\bf{27}} 324-339. 

\bibitem{Janssen} Janssen M Mecklenburg M 2006 From Classical to Relativistic Mechanics: Electromagnetic Models of the Electron \textit{Interactions: Mathematics,
Physics and Philosophy, 1860-1930} V.F. Hendricks, K.F. Jørgensen, J. Lützen and S.A. Pedersen eds. (Dodrecht: Springer) 65–134. 
\bibitem{Einstein1} Einstein A (1905) Zur Elekrodynamik bewegter K\"orper   \AP {\bf 17} 891-921.
\bibitem{Wigner} In\"on\"u  E Wigner E P 1953 On the contraction of groups and their representations  
\textit{Proc. Nat. Acad. Sci. USA} {\bf 39} 510–524.
\bibitem{Debergh} Debergh N Petit J-P D'Agostini G 2018 \textit{J. Phys. Comm.}  {\bf 2} 115012.
\bibitem{Hasenoehrl} Hasen\"ohrl F 1904  Zur Theorie der Strahlung in bewegten K\"orpern \AP {\bf 15} 344-370.
\bibitem{Hasenoehrl1} Hasen\"ohrl F 1905 Zur Theorie der Strahlung in bewegten K\"orpern: Berichtingung  \AP {\bf 16} 589–592.

\bibitem{Rothman} Boughn S Rothman T 2011 Hasen\"ohrl and the Equivalence of Mass and Energy $<$arXiv:1108.2250S$>$. 

\bibitem{Lorentz} Lorentz H A 1900 \"Uber die scheinbare Masse der Elektronen \textit{Phys. Zeitschrift} {\bf 2} 78-80.
\bibitem{Poincare2} Poincar\'e H 1906 Sur la dynamique de l'\'electron {\textit{Rendiconti del Circolo Matematico di Palermo}} {\bf {21}} 129-175.
\bibitem{Fermi} Fermi E 1922 \"Uber einen Widerspruch zwischen der elektrodynamischen und der
relativistischen Theorie der elektromagnetischen Masse {\textit{ Phys. Zeitschrift}} {\bf{23}} 340-344.  
\bibitem{Casimir} Casimir H B 1953 \textit{Physica} {\bf 19} 846.
\bibitem{Boyer} Boyer T H 1968 \PR {\bf 174} 1764. 
\bibitem{Rohrlich} Rohrlich F 1970  \textit{Am. J. Phys.} {\bf 38}, 1310–1316.
\bibitem{Moylan} Moylan P 1995 \textit{Am. J. Phys.} {\bf 63} 818-820.
\bibitem{Boughn} Boughn S 2013 Fritz Hasen\"ohrl and $E$=$mc^2$  \textit{Eur. Phys. J. H} {\bf 38} 261–278.  


\bibitem{Esposito}  Esposito A Krichevsky R and Nicolis A (2019) \PRL122, 084501.


\bibitem{Kittel} Kittel C 1971 \textit{Introduction to Solid State Physics} (New York: J. Wiley and Sons) p~196

\bibitem{Poincare3} Poincar\'e H  1895 \`A Propos de la Th\'eorie de M. Larmor \textit{L'\'eclairage \'Electrique} {\bf 5} 14. 
\bibitem{Zahar} Zahar E 2001 \textit{Poincar\'e's Philosophy} (Chicago: Open Court) pp ~113,  p~117
\bibitem{Martinez} Martinez  A A 2009 {\it { Kinematics: The Lost Origins of Einstein's Relativity}}, (Baltimore: Johns Hopkins University Press) p~256. [Martinez finds it unlikely that Einstein was not aware of Poincar\'e's Festschrift paper at the time of his {\textit{annus mirabilis}}. Specifically, he writes: ``Poincaré’s article of 1900 
\dots virtually, “was read by every expert in the field” of 
electrodynamics. \dots ``By early 1906, at the latest, Einstein was familiar with it \dots."]

\bibitem{Fadner} Fadner W L 1988 \textit{Am. J. Phys.} {\bf 56} 114


\end{thebibliography}
\end{document}